# Chiral superconductors and competing states across a Lifshitz transition in rhombohedral pentalayer graphene

Chuanqi Zheng[1,2†], Cheng Xu[3,4†], Chushan Li[1,2], Chenyu Zhang[1], Zijing Zhang[1], Kenji Watanabe[5], Takashi Taniguchi[6], Hao Yang[1,2,7], Dandan Guan[1,2,7], Liang Liu[1,2,7], Shiyong Wang[1,2,7], Yaoyi Li[1,2,7], Hao Zheng[1,2,7], Canhua Liu[1,2,7], Jinfeng Jia[1,2,7,10], Shengwei Jiang[1,2], Zhiwen Shi[1,2], Guorui Chen[1,2], Fengcheng Wu[8], Yang Zhang[3,9*], Tingxin Li[1,2,7*] and Xiaoxue Liu[1,2,7*]

[1]State Key Laboratory of Micro-nano Engineering Science, Tsung-Dao Lee Institute, Shanghai Jiao Tong University, Shanghai, China

[2]Key Laboratory of Artificial Structures and Quantum Control (Ministry of Education), School of Physics and Astronomy, Shanghai Jiao Tong University, Shanghai, China

[3]Department of Physics and Astronomy, University of Tennessee, Knoxville, TN, USA

[4]Max Planck Institute for Chemical Physics of Solids, Dresden, Germany

[5]Research Center for Electronic and Optical Materials, National Institute for Materials Science, 1-1 Namiki, Tsukuba, Japan

[6]Research Center for Materials Nanoarchitectonics, National Institute for Materials Science, 1-1 Namiki, Tsukuba, Japan

[7]Hefei National Laboratory, Hefei, China

[8]School of Physics and Technology, Wuhan University, Wuhan, China

[9]Min H. Kao Department of Electrical Engineering and Computer Science, University of Tennessee, Knoxville, Tennessee, USA

[10]Department of Physics, Southern University of Science and Technology, Shenzhen 518055, China

[†]These authors contributed equally to this work.

[*]Emails: yangzhang@utk.edu, txli89@sjtu.edu.cn, xxliu90@sjtu.edu.cn

## Abstract

Rhombohedral multilayer graphene hosts a distinctive low-energy electronic structure in which strong Coulomb interactions and nontrivial quantum geometry intertwine to generate exotic quantum states. Recent experiments reported signatures of chiral superconductivity in electron-doped rhombohedral multilayer graphene within the spin- and valley-polarized regime. Here we map the normal-state fermiology surrounding chiral superconductivity in rhombohedral pentalayer graphene. Quantum oscillation measurements reveal an electrically controlled Lifshitz transition between a simply-connected circular quarter-metal Fermi surface and an annular quarter-metal Fermi surface. The Lifshitz boundary itself shifts with perpendicular magnetic field, consistent with the strongly momentum-dependent orbital magnetic moment of the low-energy band. Approaching the transition from either side, the electron effective mass becomes strongly enhanced, implying the formation of a nearly dispersionless band bottom and a strongly reduced kinetic-energy scale. This singular electronic structure produces a regime of exceptionally strong instability in which chiral superconductivity competes with Wigner crystalline phases and reentrant quantum Hall states. In particular, two superconducting regions with signatures of orbital time-reversal-symmetry breaking lie on opposite sides of the Lifshitz boundary and have comparable transition temperatures, yet the annular-side state is suppressed by a substantially smaller perpendicular magnetic field. Our calculation finds comparable chiral pairing tendencies on the two parent Fermi surfaces while producing a much lower orbital-Zeeman pair-breaking scale and an additional finite-momentum pairing tendency for the annular state. These results identify Fermi-surface topology as a key control parameter for chiral superconductivity in rhombohedral graphene.

## Introduction

Rhombohedral multilayer graphene has emerged as a remarkably clean and electrically tunable platform for realizing strongly correlated and topological quantum matter. In an $N$-layer rhombohedral stack, the low-energy electronic states are predominantly localized on the outermost graphene layers and, in the idealized low-energy limit, acquire a strongly nonlinear dispersion that scales approximately as $E \sim k^N$. An applied perpendicular displacement electric field $D$ provides an additional powerful control knob by reshaping the low-energy bands and modifying their bandwidth, Berry curvature and orbital magnetic moment. These ingredients have generated a rich hierarchy of interaction-driven states, including half- and quarter-metals [1-4], orbital ferromagnets and Chern insulators [5-11], electronic crystals [12-14] and unconventional superconductivity [15-36].

A particularly important development was recent reports of chiral superconductivity signatures in electron-doped rhombohedral tetra-, penta- and hexalayer graphene [31,32]. These superconducting states emerge within spin- and valley-polarized regimes and exhibit magnetic switching and hysteresis under out-of-plane magnetic-field sweeps, accompanied by anomalous Hall responses in the normal state. More recent experiments, including local magnetic imaging and thermodynamic probes, further support a close relationship between superconductivity, orbital magnetism and isospin-polarized domains [33,34]. Together, these observations establish a close association between superconductivity and spontaneous orbital time-reversal-symmetry breaking, and identify rhombohedral graphene as a promising platform for topological superconductivity.

Despite this progress, a key unresolved question is why chiral superconductivity occupies only a narrow portion of the much broader spin- and valley-polarized quarter-metal regime. Isospin polarization appears to be an important prerequisite for chiral pairing, but it is evidently not sufficient. Moreover, chiral superconductivity emerges in two nearby but separate regions of the carrier density $n$-D phase space, yet how these two regions are related and what distinguishes their superconducting properties remain unclear.

Here we address these questions by systematically mapping the normal-state fermiology surrounding chiral superconductivity in rhombohedral pentalayer graphene (RPG). We show that chiral superconductivity emerges around an electrically tunable Lifshitz transition between two spin- and valley-polarized quarter metals: a nearly circular quarter metal (CQM) with simply connected Fermi contour and an annular quarter metal (AQM) whose band minimum lies at finite momentum (relative to *K*-point). At the transition the calculated band dispersion develops a characteristic 'trash-can' form [37] with a nearly dispersionless bottom and steep walls at finite momentum. Quantum oscillations directly resolve the Fermi-surface reconstruction and reveal a strong enhancement of the cyclotron effective mass from both sides of the transition. This Lifshitz-critical flat-band regime provides a natural microscopic setting for the observed proliferation of competing low-temperature orders, including chiral superconductors, Wigner crystals and reentrant quantum Hall states. We further show that the Lifshitz boundary is magnetically tunable through the momentum-dependent orbital magnetic moment, and the two chiral superconducting regions inherit different orbital-field robustness from their parent Fermi surfaces.

**Electric-field-tuned Lifshitz transition in a spin-valley-polarized RPG**

Figure 1a shows the schematic of the device, in which rhombohedral pentalayer graphene is encapsulated by hBN and equipped with dual graphite gates, allowing independent control of the $n$ and $D$ (see Methods and Extended Data Fig. 1 for device details). Figure 1b shows longitudinal resistance $R_{xx}$ as a function of $n$ and $D$, measured at zero magnetic field and $T$ = 10 mK on the electron-doped side, while the corresponding schematic phase diagram is summarized in Fig. 1c. At charge neutrality, increasing $D$ drives a transition from a layer-antiferromagnetic insulator (LAF) to a layer-polarized insulator (LPI) [3,4]. Upon electron doping, the system develops a sequence of interaction-driven, isospin-symmetry-broken metallic states. The fermiology of these states can be inferred from the measured Shubnikov–de Haas (SdH) oscillations. Tracking these oscillation frequencies as functions of $n$ and $D$ establishes the Lifshitz boundary and confirms the reconstruction of the underlying Fermi surface (Methods). At moderate $D$, a half-metallic (HM) state occupies a broad density range and evolves into a full metal (FM) at higher carrier density. With increasing $D$, a spin- and valley-polarized regime becomes dominant. Within this regime, a simply connected CQM occupies a broad region of the phase diagram, whereas an AQM is stabilized at larger $D$. As shown in Fig. 1f, the CQM region exhibits a simple oscillation pattern with a Fermi-surface degeneracy of one, whereas the AQM region shows a grid-like oscillation pattern arising from the coexistence of two sets of Fermi surfaces (see Figs. 2,3 and Extended Data Fig. 2 for further analysis establishing the Fermi-surface structures of the CQM and AQM regions). The boundary between the CQM and the AQM corresponds to a Lifshitz transition within the spin- and valley-polarized regime. Single-particle band-structure calculations (Methods) capture this directly (Fig. 1d): with increasing interlayer potential difference Δ (between the top-most and bottom-most graphene layers), the bottom of the lowest conduction band progressively flattens and eventually inverts, evolving from a single minimum at the *K*-point into a ring of minima at finite momentum around *K*-point. Near the crossover, the exceptionally flat band bottom surrounded by steeply dispersing states forms a characteristic 'trash-can'-like dispersion, marking the CQM-AQM Lifshitz transition.

Another striking signature of this transition is the strong enhancement of the cyclotron effective mass $m^*$. Extracted from the temperature dependence of the SdH amplitude using the Lifshitz–Kosevich formalism (Extended Data Figs. 3), $m^*$ increases sharply upon approaching the Lifshitz boundary from either the CQM or AQM side (Fig. 1e). Very close to the boundary, SdH oscillations become unresolvable, consistent with a strongly reduced cyclotron energy. Within experimental resolution, the data establish a strong enhancement of the band-edge effective mass on approaching the Lifshitz-critical point. This mass enhancement provides an experimental manifestation of an exceptionally flat band bottom at the CQM-AQM boundary, where the kinetic-energy scale is strongly quenched and correlation effects are correspondingly enhanced. The Lifshitz boundary is therefore not merely a geometric change of Fermi-surface topology but an electronically singular regime prone to a variety of emergent many-body states. At millikelvin temperatures, the region near the Lifshitz boundary fragments into a complex landscape of correlated ground states, as shown in Fig. 1b and 1c. These include chiral superconductivity, Wigner crystalline phases and magnetic-field-induced reentrant quantum Hall states. Their proximity suggests that the reduced kinetic-energy scale favors several competing many-body instabilities. Intriguingly, the two chiral superconducting regions lie on opposite sides of the CQM-AQM Lifshitz transition. One region, denoted CSC1, evolves from the CQM side of the phase diagram, whereas CSC2 is associated with the AQM side. At elevated temperatures, the CQM-AQM Lifshitz boundary can be more clearly identified (Fig. 1g).

**Magnetic-field control of the Lifshitz boundary**

Figure 2a-2h show maps of $R_{xx}$ and $R_{xy}$ as functions of $n$ and $D$ at selected perpendicular magnetic field $B_{\perp}$. At $B_{\perp}$ = 0.1 T (Fig. 2a, 2b), the two chiral superconducting regions, CSC1 and CSC2, remain robust, whereas another superconducting region (SC3 in Fig. 1c) is fully suppressed. The CQM-AQM Lifshitz boundary, together with the regions hosting chiral superconductivity and competing orders, shifts systematically with increasing $B_{\perp}$, with the AQM stabilized at progressively lower $D$ (Fig. 2a-2h). This observation cannot be understood as a conventional spin Zeeman effect alone. Instead, in the spin- and valley-polarized band of rhombohedral multilayer graphene with substantial Berry curvature, Bloch electrons carry a momentum-dependent orbital magnetic moment $m_{SR}$ [38]. The magnetic-field correction to the band energy is therefore approximately $\Delta E = -m_{SR} \cdot B_{\perp}$.

Crucially, $m_{SR}$ is strongly non-uniform across the band because the Berry curvature itself is strongly momentum dependent, as shown in Fig. 2i. In the relevant regime, the orbital response is relatively weak near the center of the band but becomes substantially larger at finite momentum. The applied field therefore does not simply shift the entire conduction band rigidly. Instead, it lowers $K$+$\Delta k$ states relative to those near $K$-point, continuously deforming the trash-can-like band dispersion towards a Mexican-hat-like profile and stabilizing an annular Fermi contour (Fig. 2j). Consistent with this picture, band-structure calculations show that the AQM is stabilized at progressively lower displacement fields with increasing $B_{\perp}$ (Fig. 2k,2l).

In studies of graphene superconductivity, the normal-state fermiology underlying a superconducting phase is commonly inferred from quantum oscillations measured in fixed-$D$, $n$–$B_{\perp}$ maps [15-19,39-43]. The pronounced $B_{\perp}$ dependence of the CQM-AQM Lifshitz boundary, however, calls for particular caution when applying this approach here. The Fermi surface resolved by high-field quantum oscillations need not represent the normal-state Fermi surface from which the zero-field superconductivity emerges. Nevertheless, by continuously tracking the $B_{\perp}$ evolution of the $n$-$D$ phase diagram (Fig. 2 and Extended Data Fig. 4), we can identify the normal states from which CSC1 and CSC2 emerge. Specifically, following the evolution of the normal-state features and the CQM-AQM phase boundary, rather than simply comparing the same ($n$,$D$) coordinates at different $B_\perp$, reveals that CSC1 is continuously connected to the CQM regime, whereas CSC2 is connected to the AQM regime. We therefore identify the CQM and AQM as the parent states of CSC1 and CSC2, respectively.

We also note that above $B_\perp \sim 0.5$ T, an additional CQM' phase emerges on the lower-density side of the original CQM and extends into part of the region previously occupied by CSC1. A similar phenomenon has also been reported in rhombohedral hexalayer graphene very recently [36]. Intriguingly, CSC1 is suppressed specifically where CQM' overlaps with the original CSC1 region, effectively truncating the superconducting pocket (Fig. 2c and Extended Data Fig. 11a). One possible interpretation is that CQM' reconstructs the electronic structure near the original Lifshitz-critical flat band regime, thereby disrupting the conditions favorable for chiral superconductivity. A detailed characterization of CQM' and its possible connection to the neighboring superconducting states is beyond the scope of the present work.

**Chiral superconductivity on opposite sides of the Lifshitz transition**

Having established that CSC1 and CSC2 emerge from opposite sides of the CQM-AQM transition, we next compare their superconducting properties. Figures 3a and 3b show $R_{xx}$ as a function of $n$ and $B_\perp$, measured at $D$ = 0.935 V/nm (crossing CSC1) and 1.05 V/nm (crossing CSC2) respectiviely. As discussed above, chiral superconductivity emerges in the immediate vicinity of the CQM-AQM Lifshitz transition, where the cyclotron effective mass show divergence. Consequently, quantum oscillations of the underlying normal state cannot be directly resolved in the low-$B_\perp$ regime near the superconducting regions. Quantum oscillations become visible only at higher $B_\perp$, where the $B_\perp$-induced shift of the Lifshitz boundary drives the system away from the Lifshitz-critical regime. These high-field oscillations therefore cannot be straightforwardly extrapolated to determine the zero-field parent fermiology of the superconducting states. Nevertheless, when interpreted together with the magnetic-field evolution of the $n$-$D$ phase diagram established above, the quantum oscillation patterns surrounding CSC1 and CSC2 remain consistent with their association with the CQM and AQM, respectively: on the higher-density side of CSC1, the oscillations exhibit the characteristic CQM behavior (Fig. 3a and Extended Data Figs. 2), whereas those near CSC2 display the characteristic features of an AQM (Fig. 3b and Extended Data Figs. 2).

Figures 3c and 3d show $R_{xx}$ as a function of $n$ and $T$, measured at $D$ = 0.935 V/nm and 1.05 V/nm respectiviely. Despite their association with different normal state fermiology, CSC1 and CSC2 reach comparable maximum transition temperatures $T_c$. Figures 3e shows representative $R_{xx}$-$T$ traces for CSC1 and CSC2, yielding maximum $T_c \approx 0.59$ K for CSC1 and $T_c \approx 0.53$ K for CSC2. The evolution of CSC1 and CSC2 with temperature is further illustrated by $n$-$D$ maps spanning a broader range of the phase diagram (Extended Data Fig. 5). The observed maximum $T_c$ in current device exceed those previously reported for chiral superconductivity in RPG [31]. Both CSC1 and CSC2 exhibit magnetic switching and hysteresis under $B_\perp$ sweeps, characterizing their orbital TRS-breaking nature (Extendended Data Fig. 6). Their neighboring CQM and AQM also exhibit anomalous Hall responses, with a pronounced AHE in the CQM phase and a weaker response in the AQM phase (Extended Data Fig. 7). Despite the comparable $T_c$ in CSC1 and CSC2, a striking contrast emerges in their response to $B_\perp$. As shown in Figs. 3a,3b, and 3f, CSC1 remains robust up to $B_\perp \sim 1.27$ T, whereas CSC2 is suppressed at $B_\perp \sim 0.48$ T. This contrasting response to $B_\perp$ is also evident in Fig. 2c: at $B_\perp = 1$ T, CSC2 is fully suppressed, whereas most of the CSC1 region remains superconducting. Consistently, measurements of differential ressitance $dV_{xx}/dI$ versus dc bias current $I_{dc}$ reveal comparable temperature scales but markedly different perpendicular magnetic-field scales for CSC1 and CSC2 (Extended Data Fig. 8).

These observations demonstrate that the $B_\perp$ stability of the chiral superconducting state is not determined solely by the pairing energy scale, but is strongly influenced by the fermiology of their normal state. CSC1 develops from the CQM phase, which has a single spin- and valley-polarized Fermi pocket. This pocket occupies the region of momentum space where the orbital magnetic moment of the band is small, so a applied $B_\perp$ only weakly perturbs the states from which the condensate forms (Fig. 3g). CSC2, by contrast, develops from the AQM phase: the annular Fermi contour lies at momenta where the orbital magnetic moment is large and strongly momentum dependent, so $B_\perp$ efficiently reshapes the underlying dispersion and depairs the condensate at a much smaller $B_\perp$ (Fig. 3g).

Our pairing calculations (Methods) provide a minimal microscopic account of this contrast [44,45]. Solving the linearized gap equation for the screened Coulomb interaction on the spin- and valley-polarized bands on either side of the Lifshitz transition, we obtain leading chiral *p*-wave instabilities [46,47] with comparable transition scales (~1.5 K for the CQM and ~0.9 K for the AQM). For the CQM, the maximum occurs at zero additional pair momentum [48,49] relative to the selected valley. For the AQM, the linearized instability instead has three $C_3$-related maxima at a small finite center-of-mass (Cooper-pair) momentum $Q$ with the selected valley (Extended Data Fig. 9); selecting a single-$Q$ or multi-$Q$ [50] ordered state requires higher-order terms beyond the linearized calculation. A perpendicular magnetic field enters through the orbital Zeeman coupling to the momentum-dependent orbital magnetic moment. The resulting orbital-Zeeman pair-breaking scale is about ~8.0 T for the CQM and ~1.2 T for the AQM. Because the calculation omits vortex and gradient orbital effects, these values should not be compared quantitatively with the measured critical $B_\perp$ in experiments. Nevertheless, the calculation reproduces the qualitative difference of the $B_\perp$ response between CSC1 and CSC2, and identifies momentum-space geometry as a natural origin of such difference.

## Electronic crystallization and reentrant quantum Hall states

The Lifshitz-critical regime hosts not only chiral superconductivity but also a variety of electronic crystalline states. On the low-density side, an extended high-resistance region has recently been identified as hosting Wigner crystal and metallic Wigner crystal phases [12,13]. Notably, a small high-resistance region also appears between CSC1 and CSC2, where nonlinear transport exhibits a pronounced threshold-like response suggestive of depinning (Extended Data Fig. 10). This behavior suggests that this region may also be associated with electronic crystalline order. Upon increasing $B_\perp$, chiral superconductivity is gradually suppressed, a cascade of reentrant quantum Hall (RQH) states emerges. These RQH states are confined to the vicinity of the Lifshitz boundary and evolve together with its $B_\perp$-induced shift in displacement field. Our systematic $B_\perp$-dependent mapping shows that both their locations in the *n*-*D* phase diagram and their Chern numbers evolve with applied $B_\perp$ (Fig. 4a-d, and Extended Data Fig. 4). These RQH states are also distinct from conventional RQH states in another important aspect. In conventional systems, $R_{xy}$ in a RQH state typically returns to the integer plateau associated with the nearest Landau-level filling [51-53], consistent with a picture in which electrons in the partially filled Landau level crystallize while the underlying fully filled Landau levels remain largely unaffected. Here, by contrast, the observed Hall quantization corresponds to a Chern number substantially smaller than the total Landau-level filling factor (Fig. 4b,d,e, and Extended Data Fig. 4,11), implying that a substantial fraction of electrons participates in the crystalline reconstruction. This behavior highlights the exceptionally strong correlation effects within the Lifshitz-critical flat-band region, where the interaction scale can substantially exceed the cyclotron energy and reconstruct electronic states well beyond a single partially filled Landau level [14].

Figure 4f presents representative temperature-dependent $R_{xx}$ traces for the two $C = 3$ RQH states at $B_{\perp} = 8$T. At the base temperature, each state exhibits a nearly vanishing $R_{xx}$ minimum, flanked by resistance peaks, together with a quantized $R_{xy}$ plateau at $h/3e^2$. As the temperature increases, the $R_{xx}$ minimum gradually fills in and evolves into a resistance peak, which is eventually smeared out at higher temperatures. Correspondingly, $R_{xy}$ deviates from its low-temperature quantized plateau and approaches the conventional Hall response, $R_{xy} = B/ne = h/\nu_{LL}e^2$, at elevated temperatures. We define the characteristic melting temperature $T_c^{RQH}$ as the temperature at which $R_{xx}$ $(T)$ reaches its maximum, yielding $T_c^{RQH} \sim 350$ mK and ~450-500 mK for the $C = 2$ and $C = 3$ RQH states, respectively (Extended Data Fig. 12). This non-monotonic thermal evolution and the threshold-bias response in nonlinear d$V$/d$I$ measurements (Extended Data Fig. 10) are consistent with pinned electronic crystalline order [53].

**Discussions and conclusions**

Our results provide a unified microscopic picture for the electron-doped phase diagram of rhombohedral pentalayer graphene. The key ingredient is an electrically tunable, spin- and valley-polarized conduction band with a strongly flattened minimum in the vicinity of the CQM-to-AQM Lifshitz transition. At the transition the conduction band develops an unusually flat bottom, which quenches the kinetic energy and concentrates a large density of states along the Lifshitz boundary. This renders the electronic system susceptible to multiple competing many-body instabilities. At low density the charge channel favors crystallization, whereas at relatively high density itinerant carriers condense into a chiral superconductor. Applied $B_{\perp}$ tilts this balance, suppressing superconductivity while stabilizing the crystalline family, as manifested by the emergence of RQH states. In this picture, the Wigner crystals, the reentrant quantum Hall states and the chiral superconductors are competing many-body outcomes of the same nearly dispersionless band bottom.

The observation that CSC1 and CSC2 lie on opposite sides of the Lifshitz boundary further shows that chiral pairing does not require one unique Fermi-contour topology. Instead, both CQM and AQM can support chiral superconductivity, while imprinting markedly different orbital-field responses on the resulting condensate. The Fermi-surface topology may leave an equally sharp imprint on the topological character of the pairing itself. On a simply connected, spin- and valley-polarized CQM Fermi sea, a chiral *p*-wave condensate is expected to form a topological superconductor with an odd quasiparticle Chern number, supporting a protected chiral Majorana edge mode. On an AQM Fermi sea, by contrast, the inner and outer Fermi contours contribute cancelling windings to the quasiparticle topology, so the same pairing symmetry generically yields an even Chern number [44,54] and no protected zero-energy modes. This implies that CSC1 and CSC2 would be topologically distinct superconductors despite sharing the same order-parameter symmetry, a distinction that could be probed through edge-sensitive measurements and thermal transport [55].

## Method

### Device Fabrication

Thin graphite and hexagonal boron nitride (hBN) crystals were mechanically exfoliated from bulk crystals and used as gate electrodes and dielectric layers, respectively. Rhombohedral pentalayer graphene (RPG) regions were identified by near-field infrared microscopy, and flakes with the desired stacking order were isolated using AFM-based anodic oxidation lithography. Heterostructures were assembled using a standard dry-transfer technique with a poly(bisphenol A carbonate) (PC)/polydimethylsiloxane (PDMS) stamp. The top hBN, RPG flake, bottom hBN, and bottom graphite gate were sequentially picked up and then released onto a $Si/SiO_2$ substrate to form the heterostructure. The unrelaxed regions of the RPG flakes were selected by near-field infrared microscopy and then cleaned by AFM contact-mode scanning to remove surface contamination, followed by the transfer of graphite top gates onto the heterostructures. The completed heterostructures were patterned into Hall-bar geometries using electron-beam lithography and $CHF_3/O_2$ reactive ion etching. Electrical edge contacts were fabricated by electron-beam evaporation of Cr/Au films with thicknesses of 2/80 nm.

### Transport Measurement

The carrier density $n$ and perpendicular displacement field $D$ are independently controlled by the top- and bottom-gate voltages, $V_t$ and $V_b$, respectively, and are determined as $n = (c_t V_t + c_b V_b)/e + n_0$ and $D = (c_t V_t - c_b V_b)/2\varepsilon_0 + D_0$. Here, $c_t$ and $c_b$ denote the geometric capacitances per unit area of the top and bottom gates, respectively, $\varepsilon_0$ is the vacuum permittivity, and $n_0$ and $D_0$ account for the residual doping and built-in displacement field, respectively.

Low-temperature transport measurements were carried out in a cryogen-free dilution refrigerator (Q-one) equipped with a 12 T superconducting magnet, with a base temperature of approximately 10 mK. The measurement lines were equipped with silver-epoxy filters and multistage RC filters for low-temperature noise reduction. Longitudinal and transverse resistances were measured using a standard low-frequency lock-in technique with an excitation current of 1 nA and frequencies ranging from 3.77 to 23.37 Hz. To eliminate geometric mixing between longitudinal resistance $R_{xx}$ and Hall resistance $R_{xy}$, symmetrized $R_{xx}$ and antisymmetrized $R_{xy}$ were obtained by $R_{xx}(\pm B) = [R_{xx}(+B) + R_{xx}(-B)]/2$ and $R_{xy}(\pm B) = [R_{xy}(+B) - R_{xy}(-B)]/2$, respectively.

### FFT analysis

To characterize the normal-state Fermi surfaces, we define the normalized oscillation frequency as $f_v = f_{FFT} \times e/nh$, where $f_{FFT}$ is the quantum oscillation frequency obtained from the Fourier transform of $R_{xx}(1/B_\perp)$, and $n$, $h$ and $e$ are the total carrier density, Planck's constant and electron charge, respectively. The normalized frequency $f_v$ represents the Fermi-surface area enclosed by a given cyclotron orbit, normalized by the area corresponding to the total carrier density $n$. For CSC1, the oscillations yield a single normalized frequency $f_v = 1$, consistent with a circular quarter-metal (CQM) phase. By contrast, two oscillation frequencies are resolved around CSC2, with their normalized values satisfying $|f_v^1 - f_v^2| = 1$, as expected for the inner and outer cyclotron orbits of an annular quarter-metal (AQM) Fermi surface. These results indicate that CSC1 and CSC2 emerge from distinct normal-state Fermi surfaces associated with the CQM and AQM phases, respectively.

### Single-particle continuum model

We use the $k \cdot p$ continuum model of $N$-layer rhombohedral graphene [48] in the layer $\otimes$ sublattice basis $\Psi_{\boldsymbol{\tau k}} = (A_1, B_1, \ldots, A_N, B_N)^{\mathsf{T}}$. Momenta are measured from the graphene $K$ point of the selected valley and are continuous variables truncated on a finite mesh. With $\Pi_\tau = \tau k_x + i k_y$ and $\tau = \pm 1$ for the $K$ and $K'$ valleys, the $2 \times 2$ blocks are

$$h_l^{\mathrm{intra}} = \begin{pmatrix} U_l + \delta_l & v_0 \Pi_\tau^\dagger \\ v_0 \Pi_\tau & U_l \end{pmatrix}, \quad h^{l\to l+1} = \begin{pmatrix} v_4 \Pi_\tau^\dagger & v_3 \Pi_\tau \\ \gamma_1 & v_4 \Pi_\tau^\dagger \end{pmatrix}, \quad h^{l\to l+2} = \begin{pmatrix} 0 & \gamma_2/2 \\ 0 & 0 \end{pmatrix}, \qquad (1)$$

each taken together with its Hermitian conjugate, with velocities $v_j = \frac{\sqrt{3}}{2} a_0 \gamma_j$ for $j = 0,3,4$. The layer potential is linear in the layer index, $U_l = -\Delta/2 + \Delta(l-1)/(N-1)$, with $\Delta$ is the top-to-bottom potential drop across the five layers. The surface potential $\delta$ acts on the $A_1$ and $B_N$ orbitals only, and $\gamma_2$ connects $A_l$ to $B_{l+2}$ only.

Throughout we use the parameters of Ref. [48], held fixed in every calculation reported here: intralayer nearest-neighbour hopping $\gamma_0 = 3100\,\mathrm{meV}$, vertical dimer coupling $\gamma_1 = 380\,\mathrm{meV}$, remote hopping $\gamma_2 = -15\,\mathrm{meV}$, trigonal warping $\gamma_3 = -290\,\mathrm{meV}$, interlayer skew hopping $\gamma_4 = -141\,\mathrm{meV}$, surface site energy $\delta = -10.5\,\mathrm{meV}$, lattice constant $a_0 = 0.246\,\mathrm{nm}$ and interlayer spacing $d_0 = 0.335\,\mathrm{nm}$. The hopping parameters are signed, so the corresponding velocities are $v_0 = 660.4\,\mathrm{meV\,nm}$, $v_3 = -61.8\,\mathrm{meV\,nm}$ and $v_4 = -30.0\,\mathrm{meV\,nm}$. The displacement field is parameterized internally by the total electrostatic potential drop $\Delta$ from layer 1 to layer 5.

### RPA-screened interaction, perpendicular field and pairing

The perpendicular field enters the chain at the level of the single-particle dispersion only, through the orbital Zeeman shift [38]

$$\tilde{\varepsilon}_n(\boldsymbol{k}; B) = \varepsilon_n(\boldsymbol{k}) - B\,\mu_{\mathrm{B}}\, m_n^{\mathrm{SR}}(\boldsymbol{k}), \qquad (2)$$

where $m_n^{\mathrm{SR}}$ is the self-rotation part of the orbital magnetic moment. It is evaluated from a perturbative sum over remote bands. With the velocity matrices $H_{x,\tau} = \partial_{k_x} H_\tau$ and $H_y = \partial_{k_y} H_\tau$, $M_{nm}^x = \langle u_n | H_{x,\tau} | u_m \rangle$, $M_{nm}^y = \langle u_n | H_y | u_m \rangle$ and $\mathcal{N}_{nm} = \mathrm{Im}[M_{nm}^x M_{mn}^y]$,

$$m_n^{\mathrm{SR}}(\boldsymbol{k}) = -\frac{e}{\hbar} \sum_{m \neq n} \frac{\mathcal{N}_{nm}}{\varepsilon_n - \varepsilon_m} \qquad (3)$$

[35,38], with all ten bands retained in the sum. Only the first conduction band is shifted by Eq. (2); at fixed density $\mu$ is re-solved at every field, so the field acts both by deforming the dispersion and by sliding the chemical potential.

The static Lindhard polarization of the interacting band space is

$$\Pi(\boldsymbol{q}) = -g \sum_{n,n'} \int d^2k \frac{1}{(2\pi)^2} \left|\Lambda_{n'n}(\boldsymbol{k}, \boldsymbol{q})\right|^2 \frac{f(\tilde{\xi}_{n',\boldsymbol{k}+\boldsymbol{q}}) - f(\tilde{\xi}_{n,\boldsymbol{k}})}{\tilde{\xi}_{n',\boldsymbol{k}+\boldsymbol{q}} - \tilde{\xi}_{n,\boldsymbol{k}}},$$

$$\Lambda_{n'n}(\boldsymbol{k}, \boldsymbol{q}) = \langle u_{n'}(\boldsymbol{k}+\boldsymbol{q}) | u_n(\boldsymbol{k}) \rangle, \qquad (4)$$

with $\tilde{\xi}_{n\boldsymbol{k}} = \tilde{\varepsilon}_n(\boldsymbol{k}; B) - \mu$, $f$ the Fermi function and $g$ the degeneracy factor, which is unity here since the model carries a single flavour. The form factor $\Lambda$ is the Bloch-spinor overlap obtained by projecting the density operator onto the band basis with point-like orthogonal Wannier orbitals, which discards off-site overlap and the momentum dependence of the atomic form factor. The bare interaction is the dual-gate-screened Coulomb potential of a two-dimensional layer at distance $d = 20\ \mathrm{nm}$ from a metallic gate,

$$V_0(q) = \frac{2\pi e^2}{\varepsilon_r} \frac{\tanh(qd)}{q}, \qquad (5)$$

and the screened interaction is

$$V_{\mathrm{RPA}}(q) = \frac{V_0(q)}{\epsilon_{\mathrm{RPA}}(q)}, \qquad \epsilon_{\mathrm{RPA}}(q) = 1 - V_0(q)\Pi(q). \qquad (6)$$

For a pair with total momentum $\boldsymbol{Q}$, the two partners carry $+\boldsymbol{k} + \boldsymbol{Q}/2$ and $-\boldsymbol{k} + \boldsymbol{Q}/2$, the direct Cooper channel within the first conduction band is

$$V_{\boldsymbol{Q}}^{\mathrm{pair}}(\boldsymbol{k}, \boldsymbol{k}') = V_{\mathrm{RPA}}(\boldsymbol{k} - \boldsymbol{k}') \langle u_{\boldsymbol{k}+\boldsymbol{Q}/2} | u_{\boldsymbol{k}'+\boldsymbol{Q}/2} \rangle \langle u_{-\boldsymbol{k}+\boldsymbol{Q}/2} | u_{-\boldsymbol{k}'+\boldsymbol{Q}/2} \rangle. \qquad (7)$$

The dispersions of the two partners define

$$\xi_{1,\boldsymbol{k}} = \tilde{\xi}_{\boldsymbol{k}+\boldsymbol{Q}/2}, \quad \xi_{2,\boldsymbol{k}} = \tilde{\xi}_{-\boldsymbol{k}+\boldsymbol{Q}/2}, \qquad e_1 = \frac{1}{2}(\xi_1 + \xi_2), \qquad e_0 = \frac{1}{2}(\xi_1 - \xi_2), \qquad (8)$$

$e_1$ is the mean dispersion of the pair and fixes the shell of states that can pair, while $e_0$ is the mismatch between the two partners and is what breaks them. The linearized gap equation then reads

$$\lambda(T,\boldsymbol{Q})\,\Delta_{\boldsymbol{Q}}(\boldsymbol{k}) = -\frac{1}{\Omega}\sum_{\boldsymbol{k}'} V_{\boldsymbol{Q}}^{\text{pair}}(\boldsymbol{k},\boldsymbol{k}')\;\mathcal{C}_{\boldsymbol{Q}}(\boldsymbol{k}',T)\;\Delta_{\boldsymbol{Q}}(\boldsymbol{k}') \qquad (9)$$

with:

$$\mathcal{C}_{\boldsymbol{Q}}(\boldsymbol{k},T) = \frac{\tanh[\beta(e_0+|e_1|)/2] + \tanh[\beta(-e_0+|e_1|)/2]}{4\,|e_1|}, \qquad \beta = (k_{\mathrm{B}}T)^{-1}, \qquad (10)$$

The transition temperature is defined by

$$\lambda_{\max}(T_c) = 1, \qquad (11)$$

and the corresponding eigenvector is the order parameter.

## Acknowledgement

We acknowledge Long Ju, Yingming Xie and Qingdong Jiang for helpful discussions. This work is supported by the National Key R&D Program of China (Nos. 2022YFA1402702, 2022YFA1405400, 2022YFA1402404, 2021YFA1401400, 2021YFA1400100, 2020YFA0309000, 2022YFA1402401, 2020YFA0309000), the National Natural Science Foundation of China (Nos. 12350403, 92565302, 12374045, 12174250, 12141404, 12350005, 12488101), the Quantum Science and Technology-National Science and Technology Major Project (Nos. 2021ZD0302600 and 2021ZD0302500), Natural Science Foundation of Shanghai (No. 24QA2703700), and Shanghai Science and Technology Innovation Action Plan (grant No. 24LZ1401100). X. Liu, T. L. acknowledge the Shanghai Jiao Tong University 2030 Initiative Program. X. Liu acknowledges 'Shuguang Program' supported by Shanghai Education Development Foundation. T.L. acknowledges support from Asian Young Scientist Fellowship (AYSF) and the New Cornerstone Science Foundation through the XPLORER PRIZE. K. W. and T. T. acknowledge support from the JSPS KAKENHI (Nos. 21H05233 and 23H02052) and World Premier International Research Center Initiative (WPI), MEXT, Japan.

## Competing interests

The authors declare no competing financial interests.

## Data availability

All data that support the findings of this study are available from the contact author upon request.

# Figures

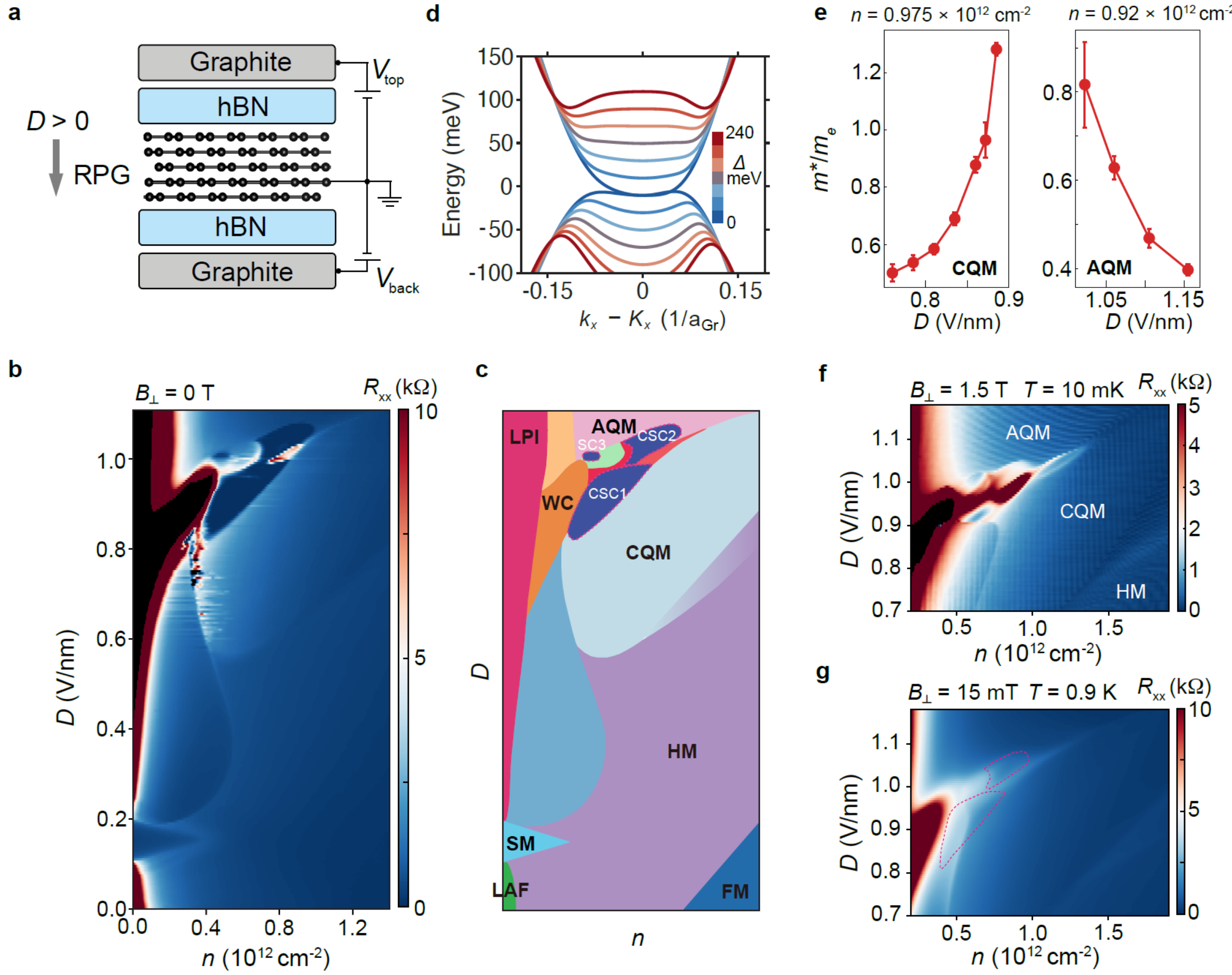

**Fig. 1 | Transport phase diagram of moiréless RPG. a**, Schematic of the dual-graphite-gated RPG device. **b**, Longitudinal resistance $R_{xx}$ as a function of carrier density $n$ and displacement field $D$, measured at $B_\perp = 0$ T and $T = 10$ mK. **c**, Schematic phase diagram corresponding to **b**, showing different spontaneous isospin-symmetry-broken states. LAF, SM, LPI, WC, AQM, CQM, HM and FM denote the layer-antiferromagnetic state, semimetallic state, layer-polarized insulator, Wigner crystal, annular quarter metal, circular quarter metal, half metal and full metal, respectively. CSC1, CSC2 and SC3 denote three superconducting states. **d**, Calculated highest valence and lowest conduction bands of rhombohedral pentalayer graphene along $k_x$ through the $K$-point, for the top-to-bottom potential drop $\Delta$ = 0-240 meV (colour scale); $a_{Gr}$ = 0.246 nm is the graphene lattice constant. **e**, Effective mass $m^*/m_e$ as a function of $D$ at fixed densities in the CQM and AQM regimes, with the corresponding $n$ values indicated in each panel. **f**, **g**, $R_{xx}$ maps as functions of $n$ and $D$, measured at $B_\perp = 1.5$ T, $T = 10$ mK (**f**) and at $B_\perp = 15$ mT, $T = 0.9$ K (**g**). The AQM, CQM and HM regions are marked in **f**. The dashed lines in **g** mark the positions of the chiral superconducting regions (CSC1 and CSC2) at $B_\perp = 15$ mT and $T = 10$ mK.

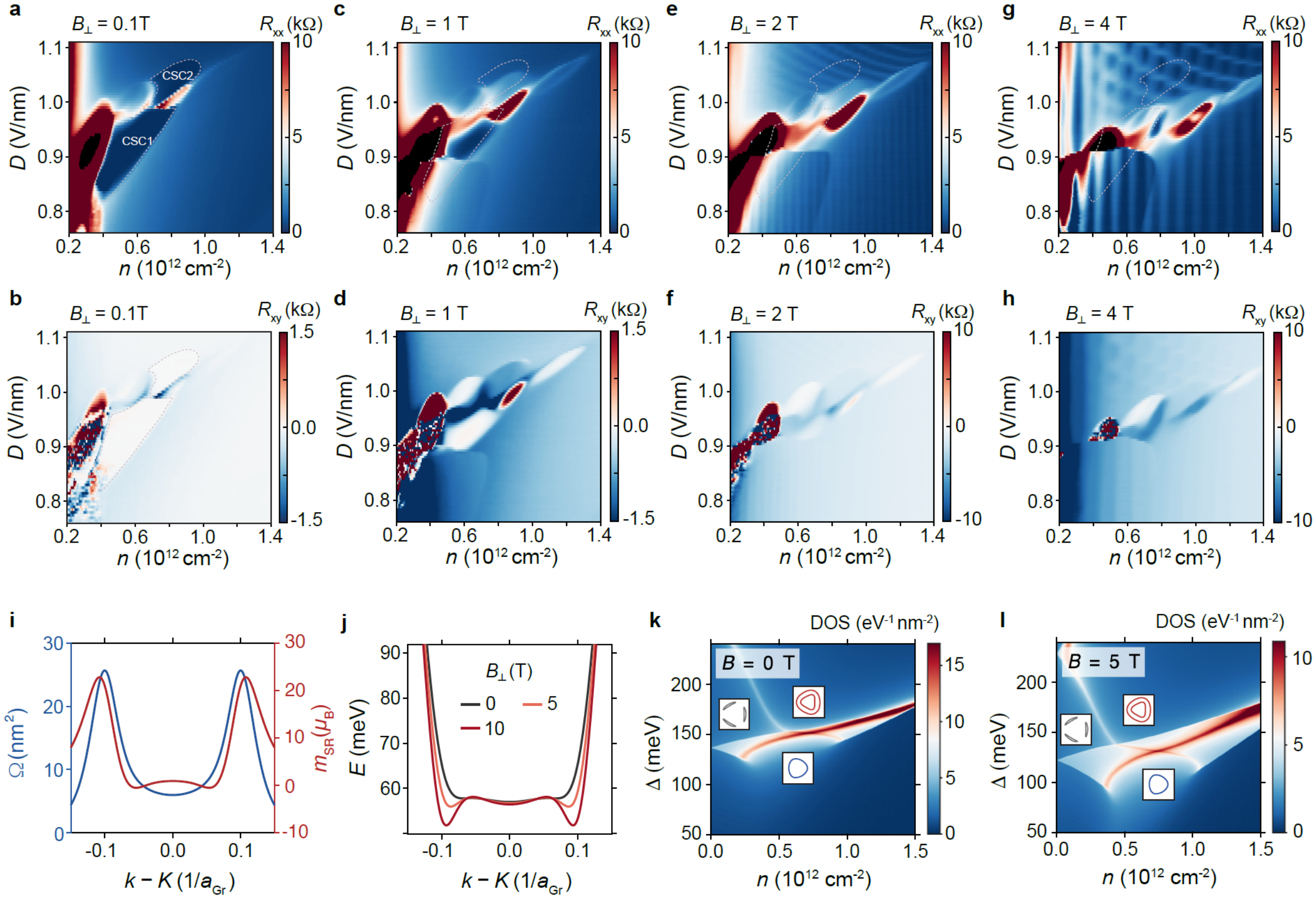

**Fig. 2 | Perpendicular magnetic-field evolution of the *n-D* phase diagram. a-h**, Symmetrized longitudinal resistance $R_{xx}$ (**a**,**c**,**e**,**g**) and anti-symmetrized Hall resistance $R_{xy}$ (**b**,**d**,**f**,**h**) as functions of carrier density $n$ and displacement field $D$, measured at $T$ = 10 mK and different perpendicular magnetic fields $B_{\perp}$, as indicated in each panel. The contours of CSC1 and CSC2, extracted from the $B_{\perp}$ = 0.1 T map, are outlined by gray dashed lines and overlaid on the $R_{xx}$ maps at $B_{\perp}$ = 1 T (**c**), 2 T (**e**), and 4 T (**g**). **i**, Calculated Berry curvature Ω (blue) and self-rotation orbital magnetic moment $m_{SR}$ (red) of the lowest conduction band in RPG at Δ = 135 meV, along a straight cut through $K$ oriented 30° from $k_x$; $k$ is measured from $K$-point. **j**, Calculated lowest conduction band dispersion at Δ = 135 meV in RPG under $B_{\perp}$ = 0, 5 and 10 T. **k**, Calculated density of states at the Fermi level in the ($n$, Δ) plane at $B$ = 0, including the orbital Zeeman shift; Δ is the top-to-bottom potential drop. **l**, Same as **k** at $B_{\perp}$ = 5 T.

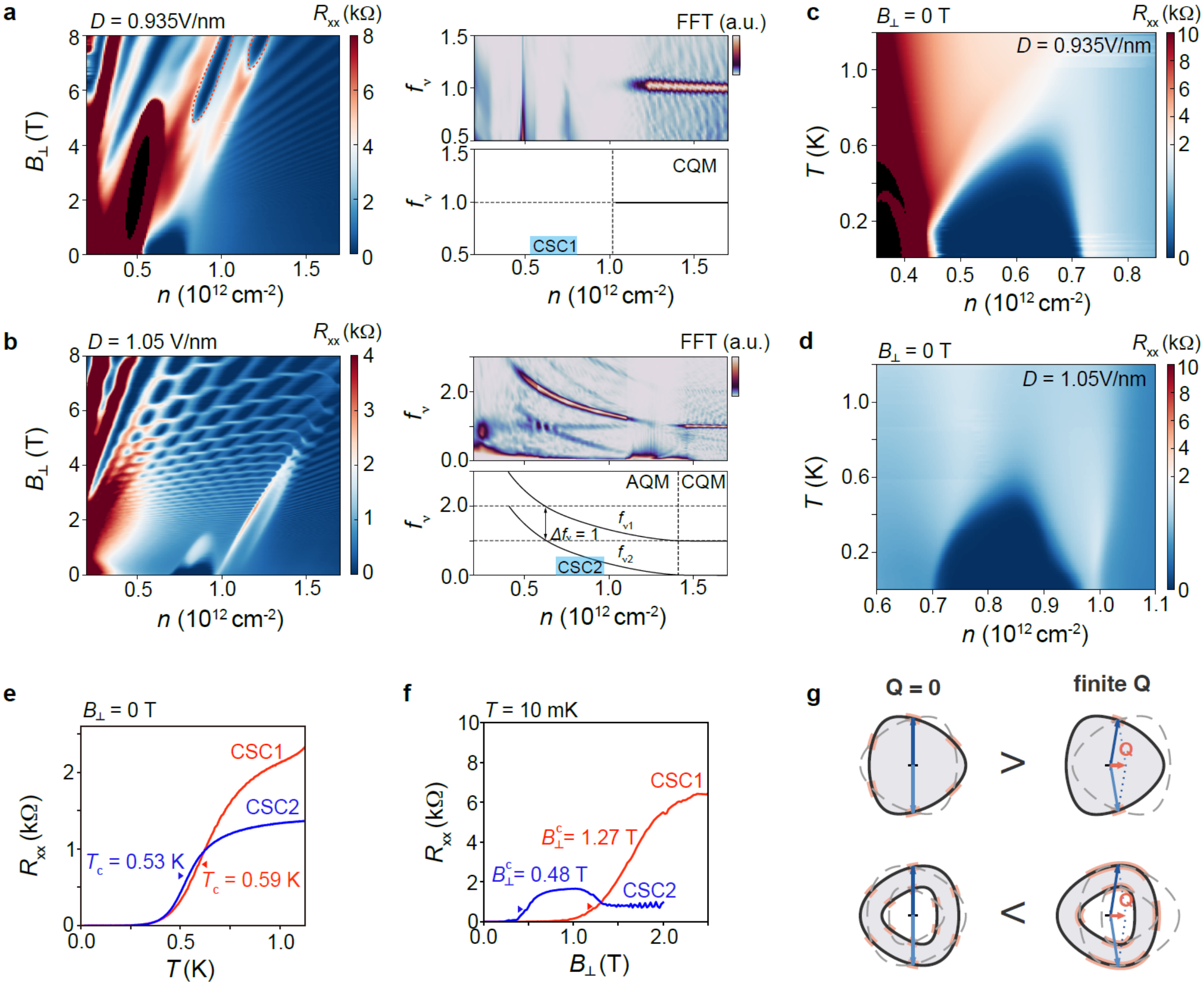

**Fig. 3 | Fermiology-dependent critical magnetic fields of CSC1 and CSC2. a**,**b**, $R_{xx}$ maps as functions of $n$ and $B_\perp$ measured at $T = 10$ mK, and fixed displacement fields $D = 0.935$ V/nm (**a**) and $D = 1.05$ V/nm (**b**). The corresponding normalized FFT spectra, calculated over $B_\perp = 1.5$ T - 8 T for **a** and 1.5 T - 6 T for **b**, are shown alongside each $R_{xx}$-$n$-$B_\perp$ map. The red dashed lines in **a** denote the RQH states. **c**, **d**, $R_{xx}$ maps as functions of $n$ and $T$, measured at $B_\perp = 0$ T, and fixed displacement fields $D = 0.935$ V/nm (**c**) and $D = 1.05$ V/nm (**d**), respectively. **e**, $R_{xx}$ as a function of $T$ at $B_\perp = 0$ T for representative points in CSC1 ($n = 0.65 \times 10^{12}$ cm$^{-2}$, $D = 0.935$ V/nm) and CSC2 ($n = 0.87 \times 10^{12}$ cm-2, $D = 1.05$ V/nm), showing that the two chiral superconducting domes have comparable maximum transition temperatures $T_c$. **f**, $R_{xx}$ as a function of $B_\perp$ at $T = 10$ mK for the same points as in **e**, showing markedly different critical perpendicular magnetic fields $B_c$ for CSC1 and CSC2. **g**, Pairing geometry at zero and finite pairing momentum $Q$, for an CQM Fermi surface (top) and the AQM Fermi surface (bottom). Solid lines: Fermi contour; dashed lines: its image under $k \to Q - k$; the blue chord connects one pair of partners. Orange arcs mark resonant sections where both partners lie on the Fermi surface. For the CQM Fermi surface the resonant phase space is largest at $Q = 0$; for the AQM Fermi surface a finite $Q$ is favoured.

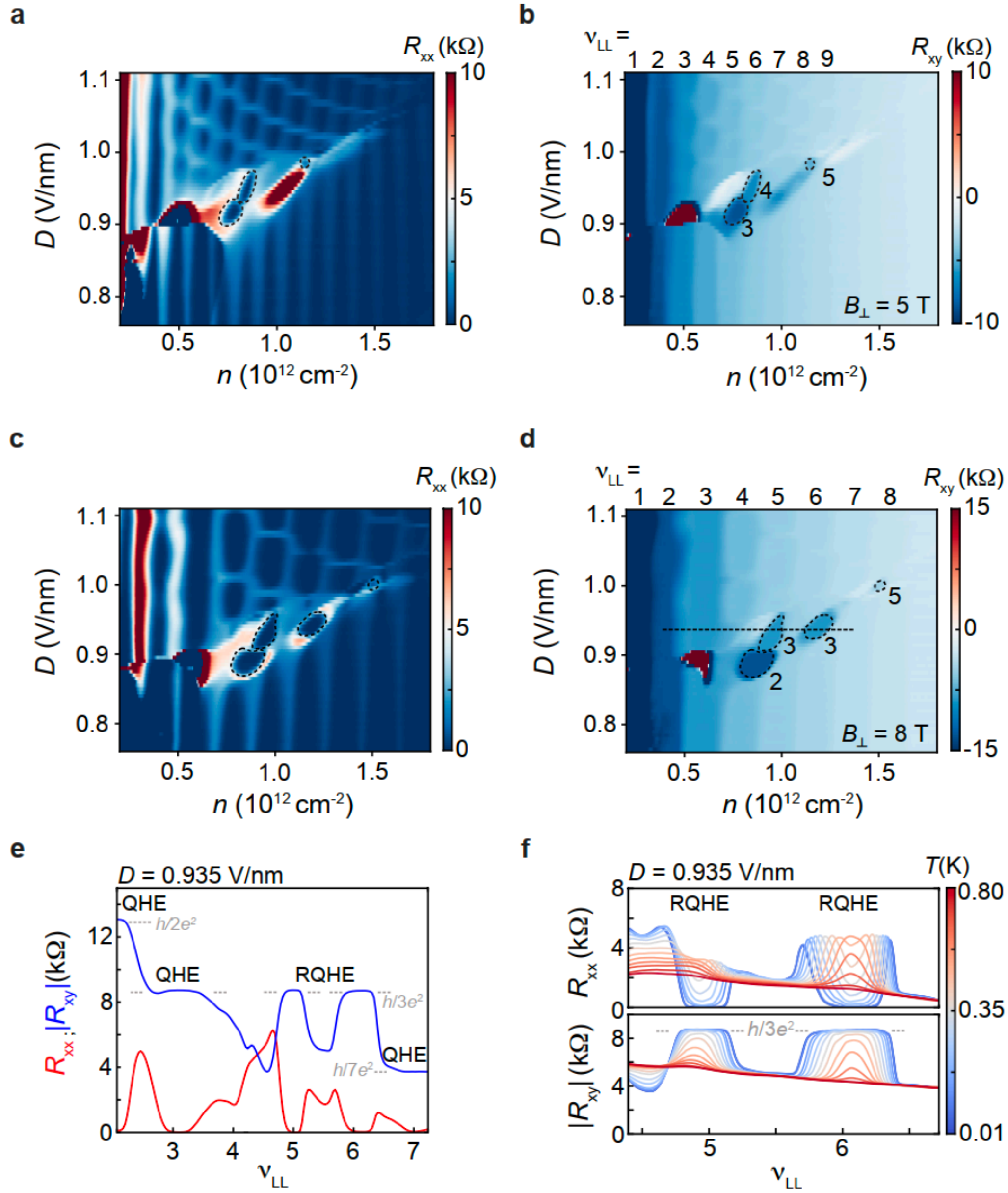

**Fig. 4 | Cascade of RQH states near the AQM-CQM Lifshitz boundary. a-d**, $R_{xx}$ (**a**,**c**) and $R_{xy}$ (**b**,**d**) maps as functions of $n$ and $D$, measured at $B_\perp$ = 5 T (**a**,**b**) and 8 T (**c**,**d**). Dashed contours highlight the RQH states, with the corresponding Chern numbers indicated in the $R_{xy}$ maps. Landau-level filling factors $\nu_{LL}$ are shown along the top axes of the $R_{xy}$ maps in **b** and **d**. **e**, $R_{xx}$ and $|R_{xy}|$ as a function of $\nu_{LL}$ at $D$ = 0.935 V/nm, corresponding to the dashed line in **d**. The QHE and RQH states are labeled. **f**, Temperature dependence of $R_{xx}$ and $|R_{xy}|$ near the $C$ = 3 RQH states.

## Extended Data Figures

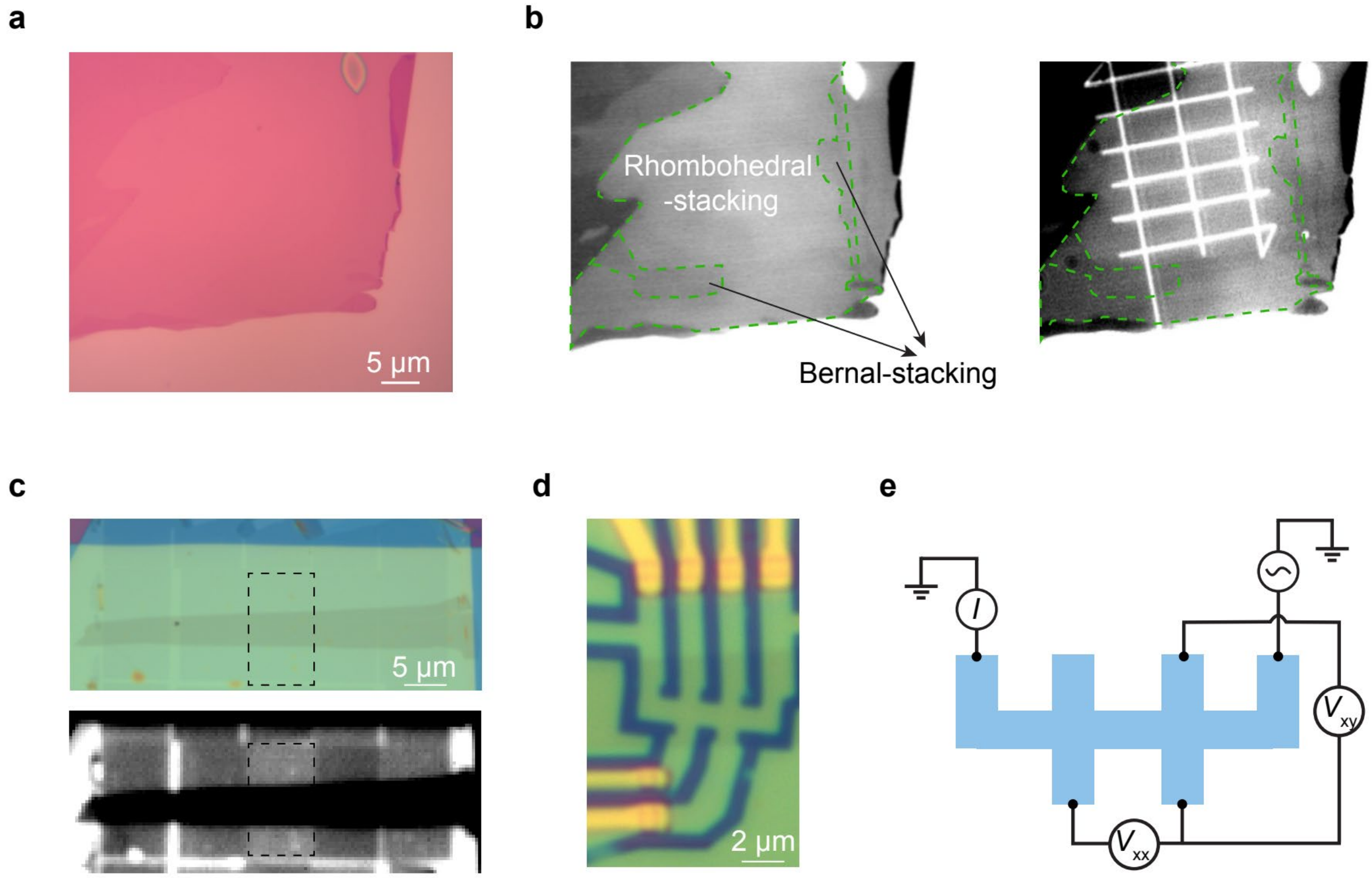

**Extended Data Fig. 1 | Device information. a**, Optical image of the RPG flake. **b**, Infrared image of the same RPG flake (left panel), where rhombohedral-stacked and Bernal-stacked regionsare shown as bright and dark areas, respectively. To improve fabrication yield, the RPG flake was cut into multiple pieces prior to the pick-up process (right panel) by the AFM lithography. **c**, Optical (upper panel) and infrared (lower panel) images of the assembled heterostructure, with the RPG flake used for device fabrication highlighted by the black dashed line. **d**, Optical image of the fabricated Hall-bar device. **e**, Schematic illustration of the measurement configuration. Scale bars are shown in each panel.

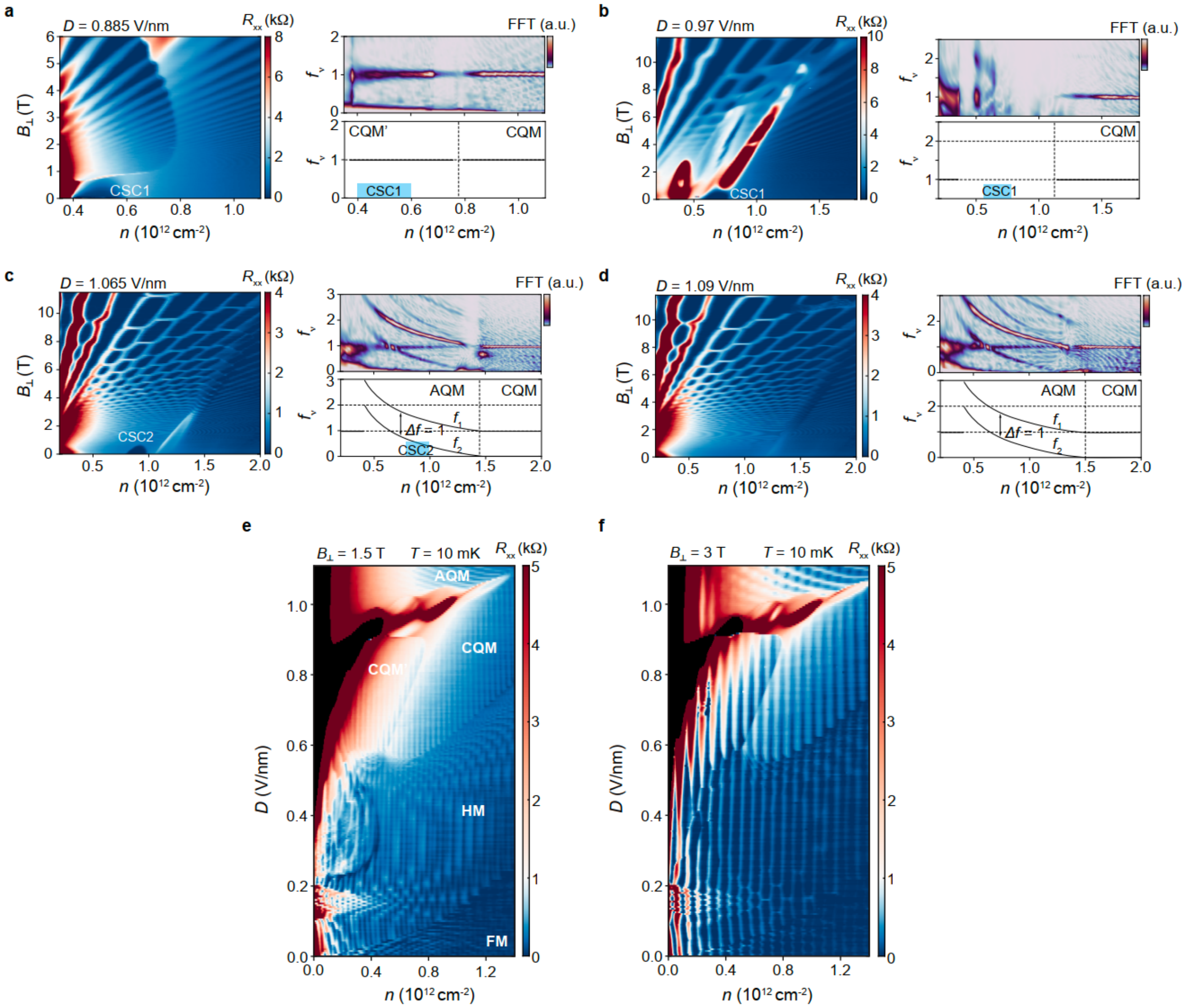

**Extended Data Fig. 2 | Landau fan diagrams and FFT analysis at different displacement fields. a-d,** Landau fan diagrams of $R_{xx}$ as functions of carrier density $n$ and perpendicular magnetic field $B_\perp$, measured at the fixed displacement fields indicated in each panel. The corresponding normalized FFT spectra are shown alongside each Landau fan diagram and are calculated over $B_\perp$ = 1.3 - 6 T (a)**,** 2 - 11.8 T (b), 1.5 - 11.5 T (c), and 1.7 - 11.8 T (d). The corresponding fermiology inferred from the FFT analysis is indicated in each panel. The blue markers in each FFT map indicate the density range of the corresponding superconducting dome at zero magnetic field. At the fixed displacement field $D$ = 0.97 V/nm in panel b, the field-induced shift of the Lifshitz boundary towards lower $D$ drives the low-density region (about 0.5-1.2 × $10^{12}$ cm$^{-2}$) into the AQM regime at high $B_\perp$. **e**, **f**, $n$-$D$ phase diagrams over a broad range of carrier density $n$ and displacement field $D$, measured at $B_\perp$ = 1.5 T (e) and 3 T (f), respectively, at $T$ = 10 mK. The corresponding isospin-symmetry-broken states are labelled in **e**.

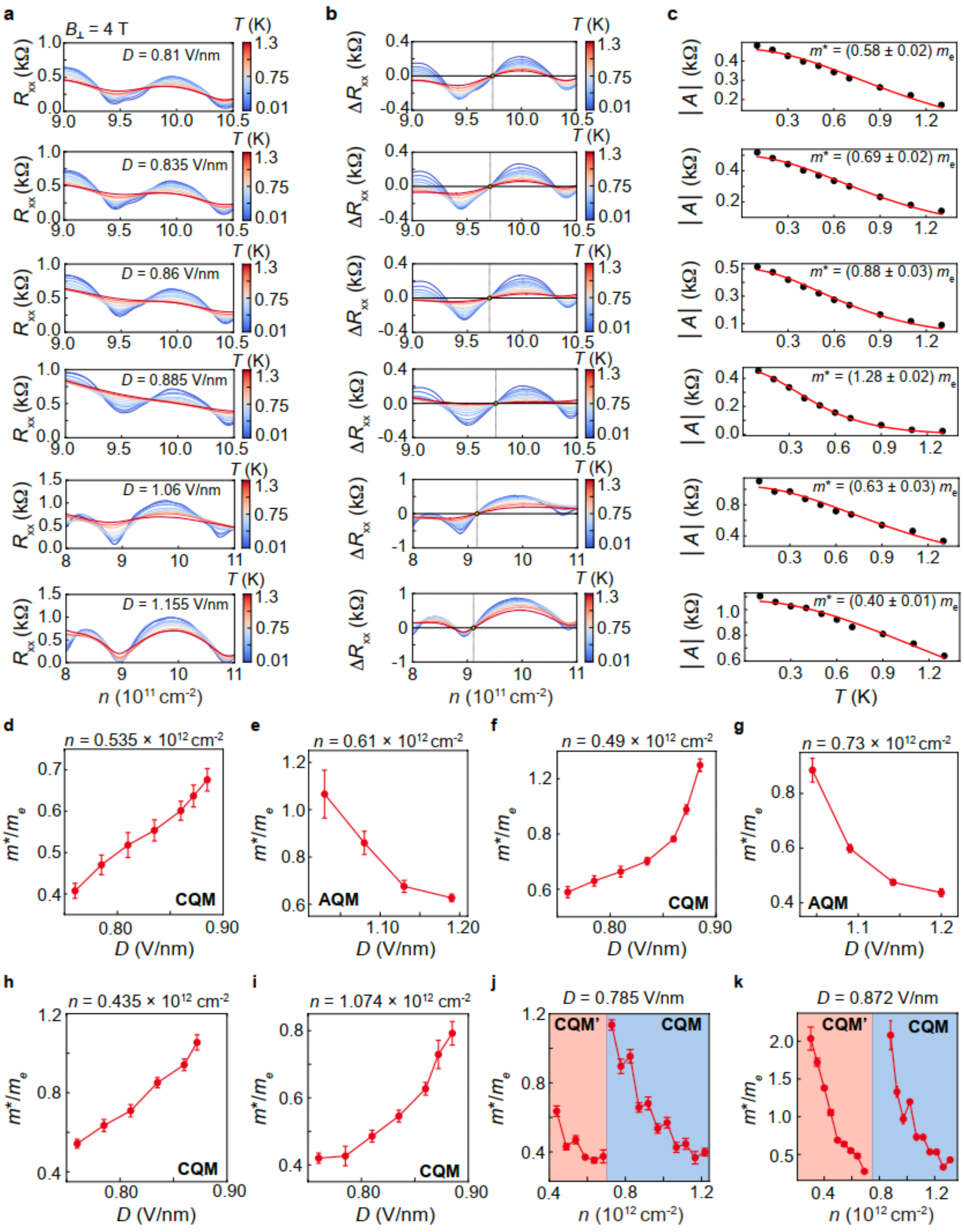

**Extended Data Fig. 3 | Effective mass extracted from temperature-dependent quantum oscillations. a-c**, Data and analysis used to extract the effective mass plotted in Fig. 1e. **a,** $R_{xx}$ as a function of carrier density $n$ measured at different displacement fields $D$, with $B_\perp = 4$ T and $T$ ranging from 10 mK to 1.3 K. **b,** Oscillatory component $\Delta R_{xx}$ of $R_{xx}$, obtained as $\Delta R_{xx} = R_{xx}$ -

$R_{xx}^{4\text{ K_shift}}$, where $R_{xx}^{4\text{ K_shift}}$ denotes the $R_{xx}$ trace at $T = 4$ K vertically shifted to the crossing point of all traces before subtraction. **c,** Temperature dependence of the oscillation amplitude $|A|$, defined as the peak-to-valley difference for each oscillation marked in b. The yellow markers in b indicate the representative densities assigned to the extracted $m^*$, located between the peak and valley used to determine $A$. Solid curves are Lifshitz-Kosevich fits used to extract $m^*$. **d-k,** Additional effective-mass data extracted using the same procedure. **d-i,** Extracted $m^*/m_e$ as a function of $D$ at fixed $n$, with the corresponding $n$ values indicated in each panel. The effective mass exhibits a pronounced enhancement and tends toward divergence upon approaching the Lifshitz transition boundary. **j, k,** Extracted $m^*/m_e$ as a function of $n$ at fixed $D$, with the corresponding $D$ values indicated in each panel. Shaded regions indicate the CQM' and CQM regimes. The effective mass exhibits an abrupt change across the boundary between the CQM' and CQM phases.

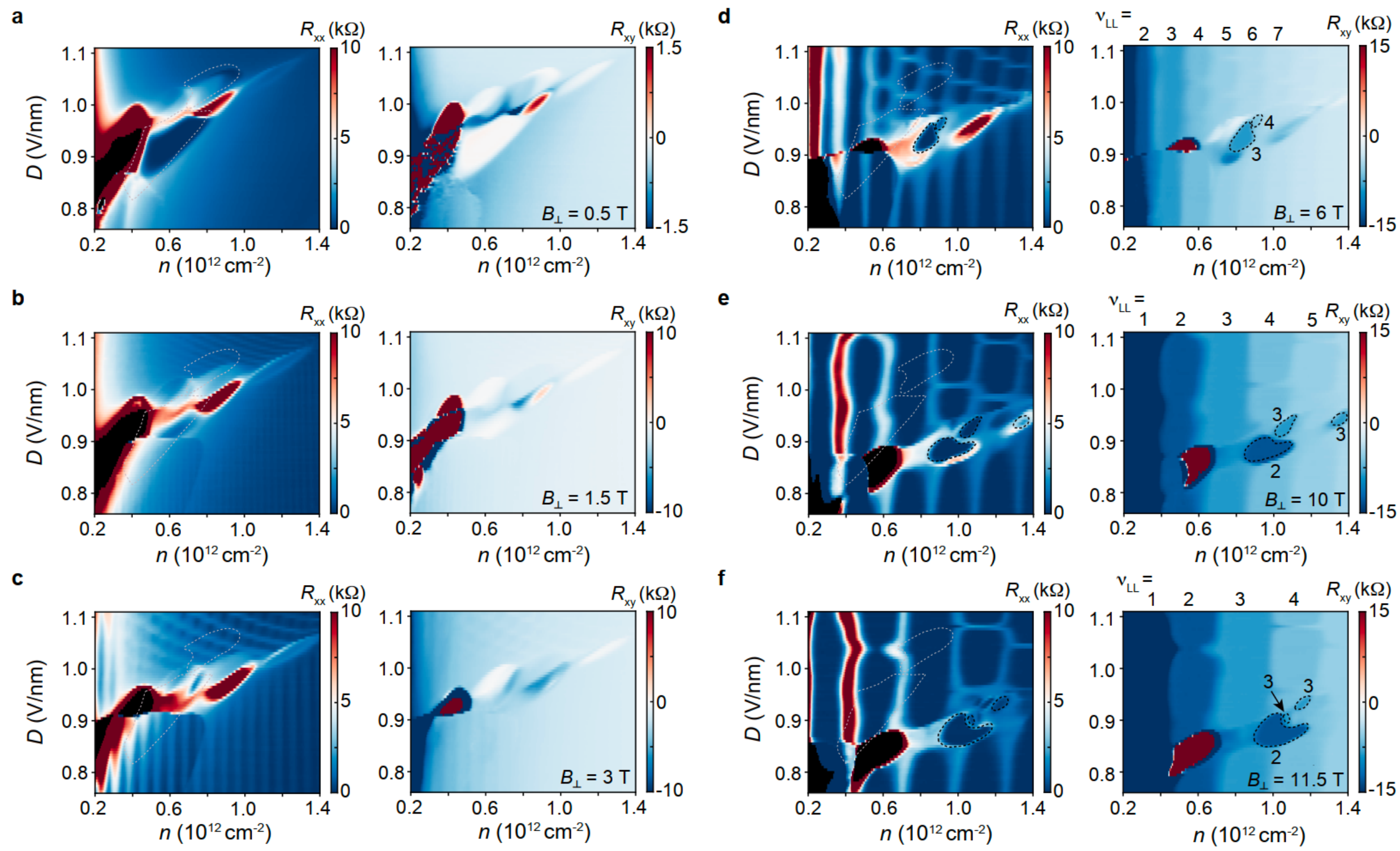

**Extended Data Fig. 4 | Additional $R_{xx}$-$n$-$D$ and $R_{xy}$-$n$-$D$ maps under perpendicular magnetic fields. a-d**, $R_{xx}$ and $R_{xy}$ maps as functions of carrier density $n$ and displacement field $D$, measured at fixed $B_{\perp} = 0.5$ T (a), 1.5 T (b), 3 T (c), 6 T (d), 10 T (e) and 11.5 T (f). Black Dashed contours highlight the RQH states, with the corresponding Chern numbers labeled in the $R_{xy}$ maps. Gray dashed contours the CSC1 and CSC2 regions extracted from the $B_{\perp} = 0.1$ T map and overlaid on the $R_{xx}$ maps. Landau-level filling factors $\nu_{LL}$ are indicated along the top axes of the $R_{xy}$ maps in **d-f**.

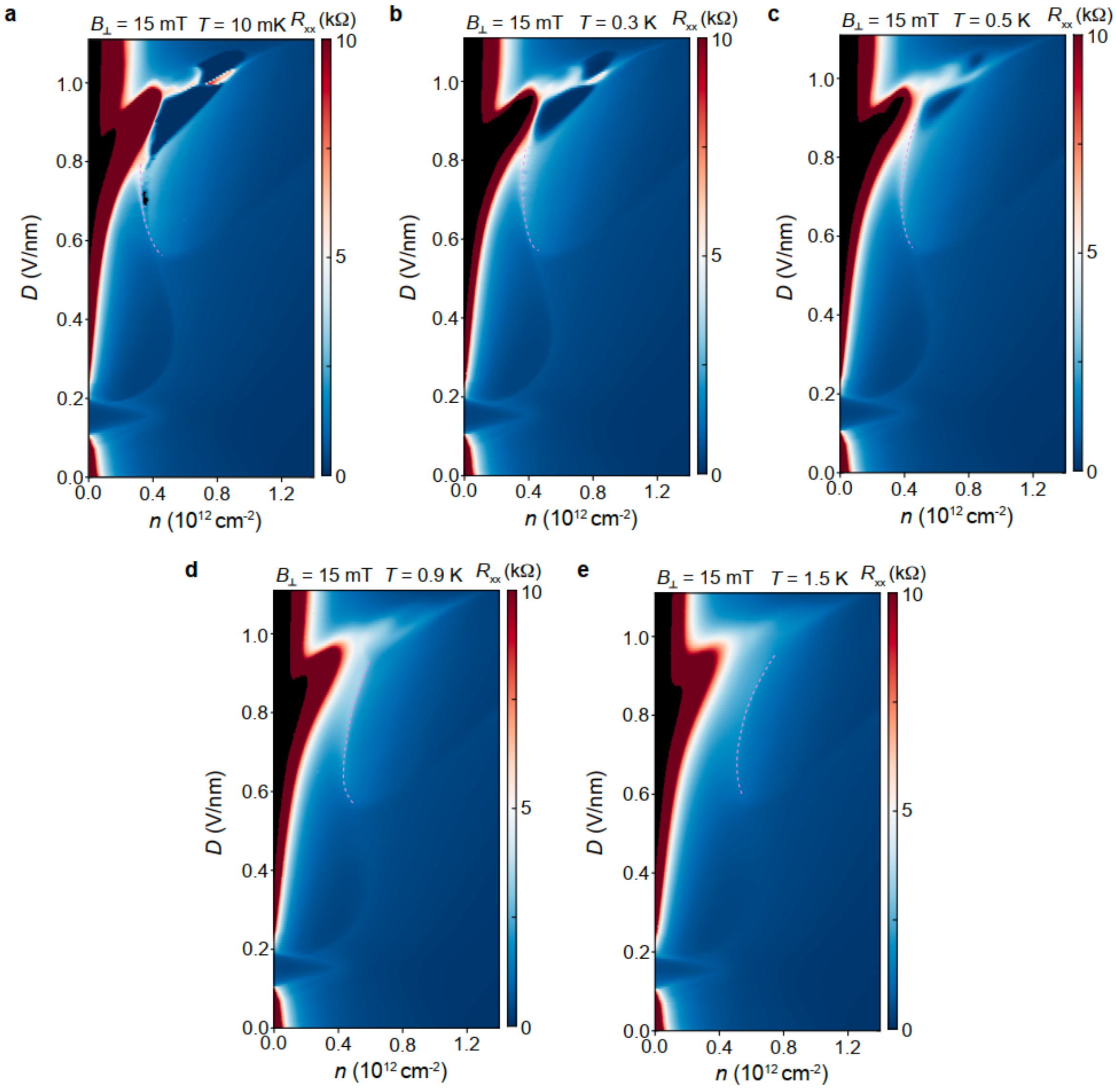

**Extended Data Fig. 5 | Temperature dependence of the *n*-*D* phase diagram. a-e**, $R_{xx}$ maps as functions of carrier density $n$ and displacement field $D$, measured at $B_\perp$ = 15 mT for $T$ = 10 mK (a), 0.3 K (b), 0.5 K (c), 0.9 K (d) and 1.5 K (e). The dashed line marks the boundary between the CQM phase and an undetermined low-density metallic state. This boundary shifts to higher density with increasing temperature, whereas the other boundaries between distinct isospin-symmetry-broken states remain nearly unchanged.

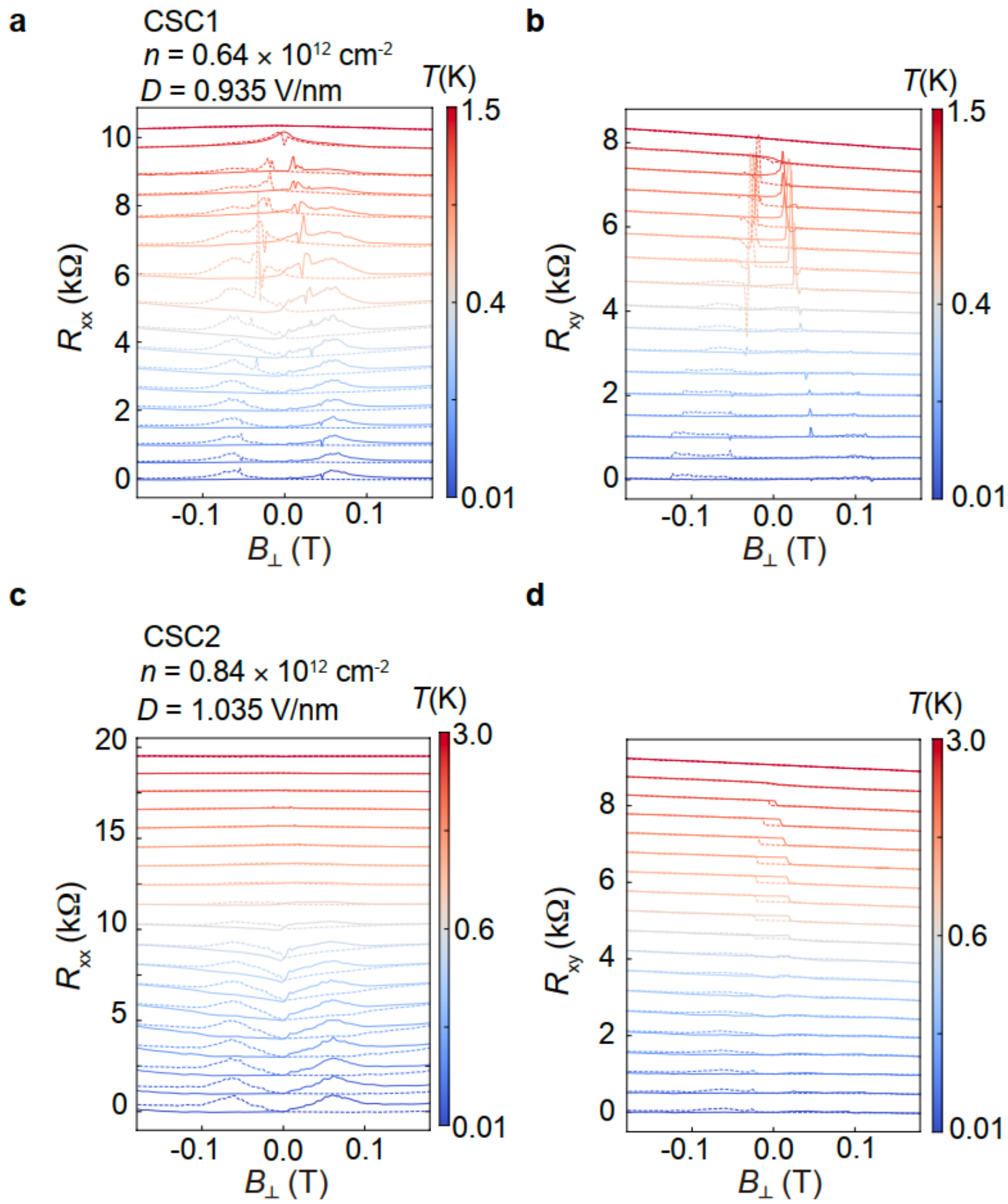

**Extended Data Fig. 6 | Magnetic hysteresis in CSC1 and CSC2**. **a**,**b**, $R_{xx}$ (a) and $R_{xy}$ (b) hysteresis loops of CSC1 versus $B_{\perp}$ at different temperatures, measured at $n = 0.64 \times 10^{12}$ cm$^{-2}$ and $D$ = 0.935 V/nm. **c**,**d**, $R_{xx}$ (c) and $R_{xy}$ (d) hysteresis loops of CSC2 versus $B_{\perp}$ at different temperatures, measured at $n = 0.84 \times 10^{12}$ cm$^{-2}$ and $D$ = 1.035 V/nm. Traces are vertically offset for clarity.

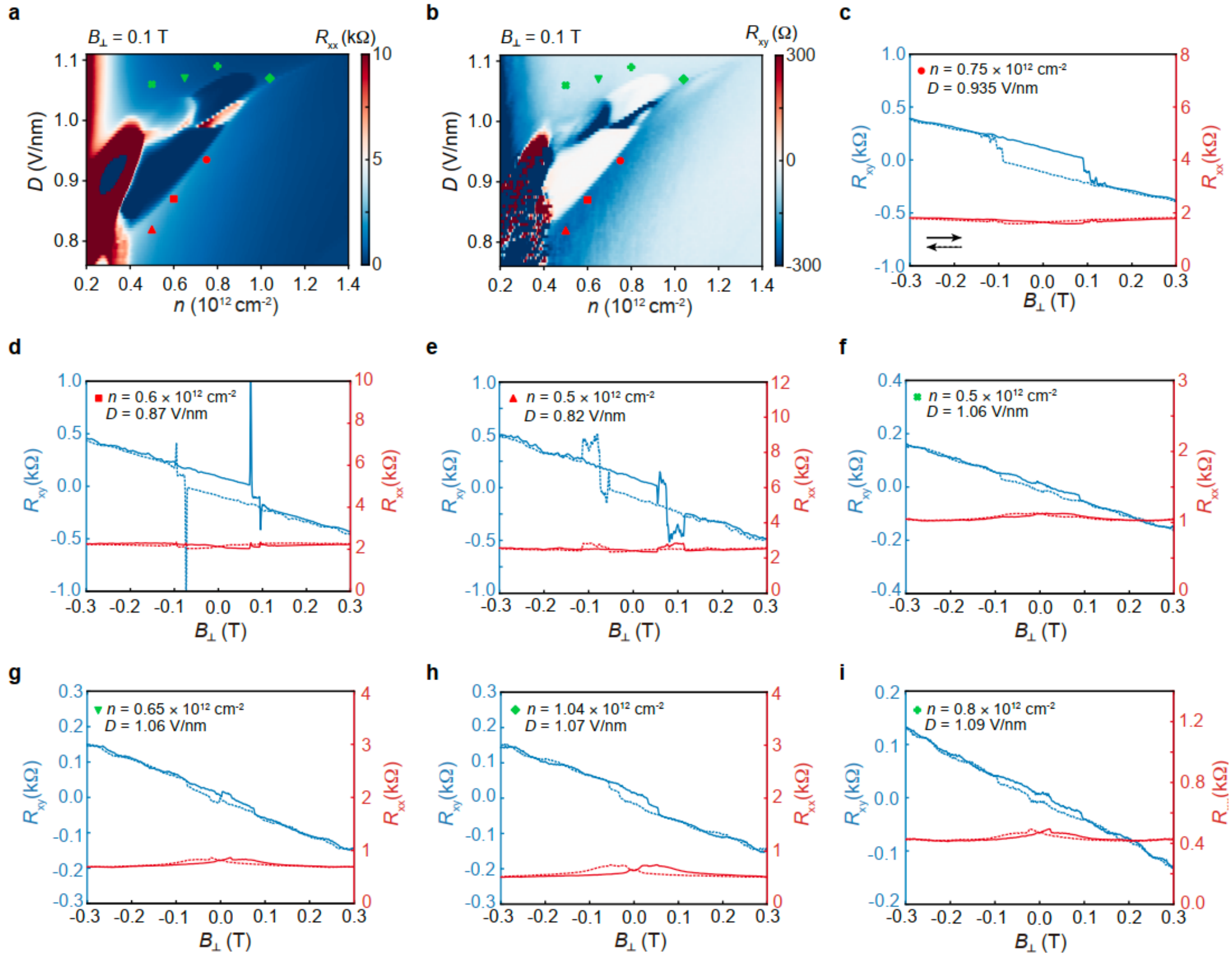

**Extended Data Fig. 7 |Anomalous Hall effect in the CQM and AQM regions. a**,**b**, Phase diagrams of $R_{xx}$ (a) and $R_{xy}$ (b) as functions of $n$ and displacement field $D$, measured at $B_\perp = 0.1$ T and $T = 10$ mK. **c**-**i**, $R_{xx}$ and $R_{xy}$ magnetic hysteresis loops measured by sweeping $B_\perp$ at selected ($n$, $D$) points marked in **a** and **b**. Red markers denote representative points in the CQM regime, whereas green markers denote those in the AQM regime. An anomalous Hall response is observed in both the CQM and AQM regimes, with a substantially stronger response in the CQM regime.

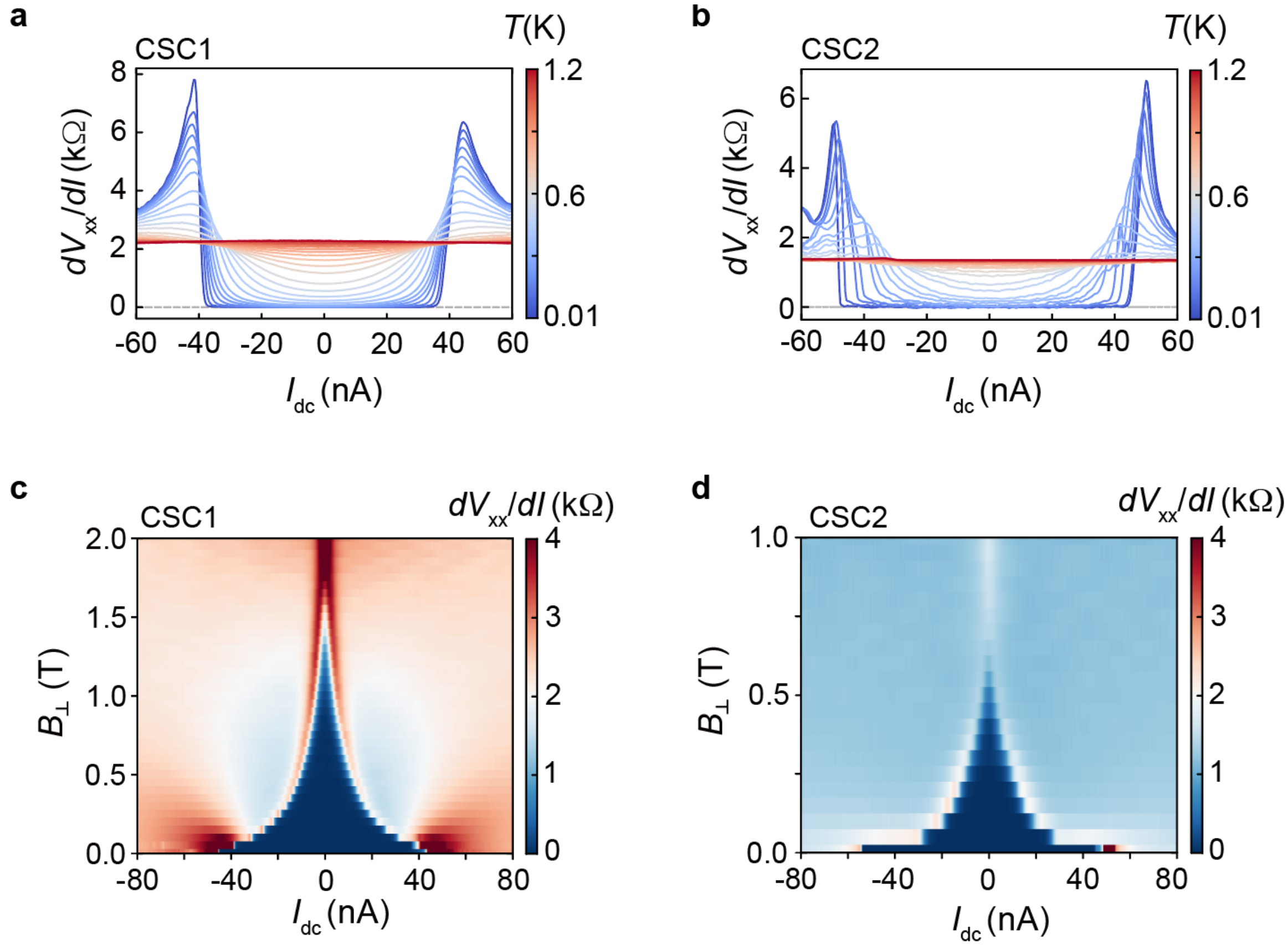

**Extended Data Fig. 8 | Differential resistance measurements of CSC1 and CSC2. a**, **b**, Temperature dependence of d$V_{xx}$/d$I$ as a function of $I_{dc}$ for representative points in CSC1 ($n$ = 0.65 × $10^{12}$ cm$^{-2}$, $D$ = 0.935 V/nm, $B_\perp$ = 50 mT) (a) and CSC2 ($n$ = 0.87 × $10^{12}$ cm-2, $D$ = 1.05 V/nm, $B$ = 0) (b). **c**,**d**, d$V_{xx}$/d$I$ as functions of $I_{dc}$ and $B_\perp$ *at* $T$ = 10 mK, measured at the same points as in **a** and **b**, respectively. The nonlinear d$V_{xx}$/d$I$ response is suppressed at comparable temperatures, whereas the perpendicular magnetic-field for suppressing superconductivity in CSC1 is nearly three times larger than that of CSC2.

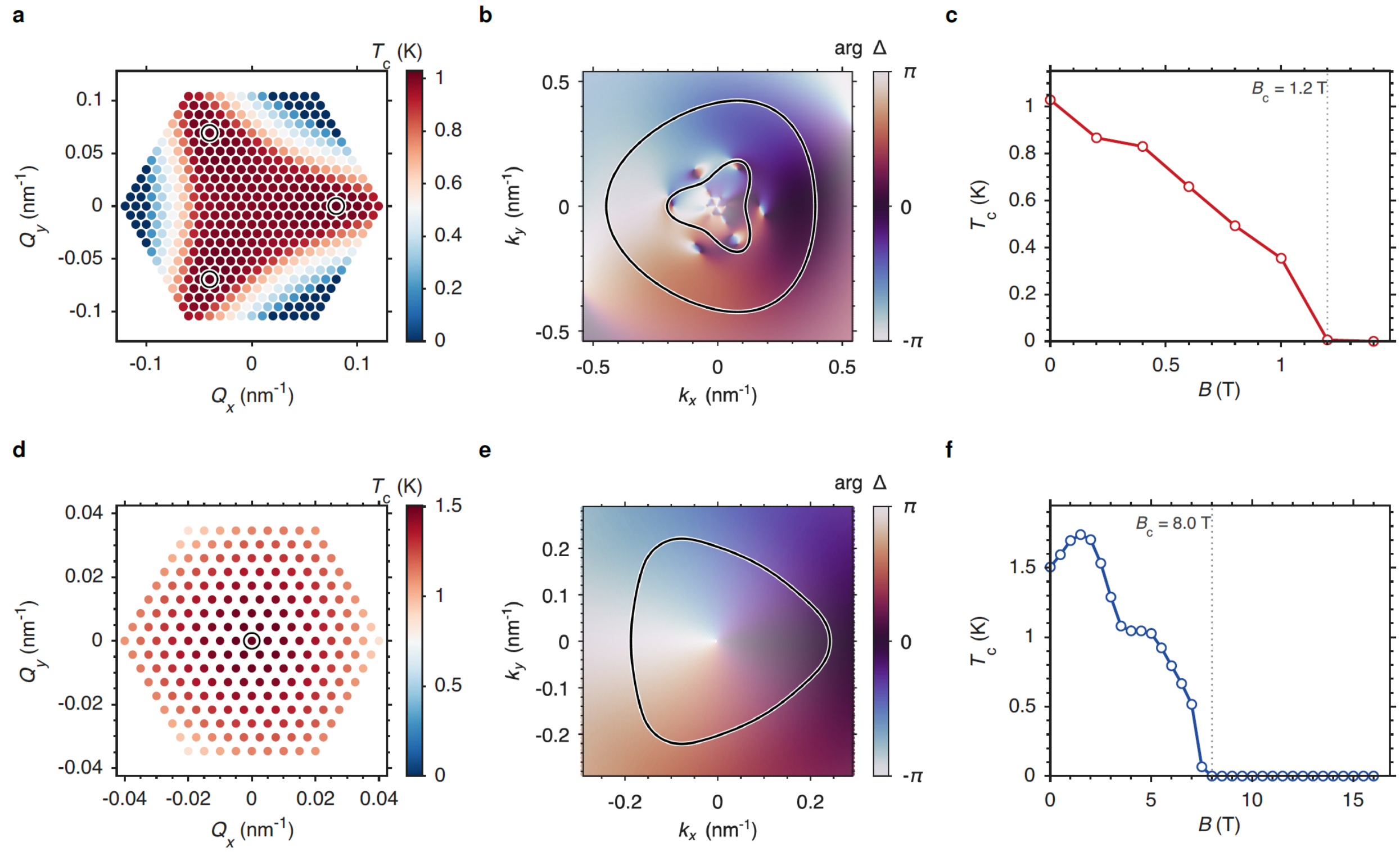

**Extended Data Fig. 9 | AQM versus CQM: pairing momentum, order parameter and field resilience.** Calculated superconducting properties at two representative points, with an annular Fermi contour (**a**–**c**; $n$ = 1.20×10$^{12}$ cm$^{-2}$, Δ = 170 meV) and a circular Fermi contour (**d**–**f**; $n$ = 0.35×10$^{12}$ cm$^{-2}$, Δ = 115 meV), for the same dielectric environment ($\varepsilon_r$ = 6). **a,d,** Transition temperature $T_c$ versus pairing momentum $Q$; circles mark the maxima — three $C_3$-equivalent finite-$Q$ peaks at $|Q|$ = 0.08 nm$^{-1}$ for the annulus, and $Q$ = 0 for the circular. **b,e,** Order parameter at the optimal $Q$ and $B$ = 0: colour, arg $\Delta_{SC}(k)$; black line, Fermi surface. Both states have winding number $l$ = −1. **c,f,** $T_c$ versus perpendicular field $B_\perp$ (orbital Zeeman coupling), with $B_c$ marked: 1.2 T for the finite-$Q$ annulus state against 8.0 T for the circular case.

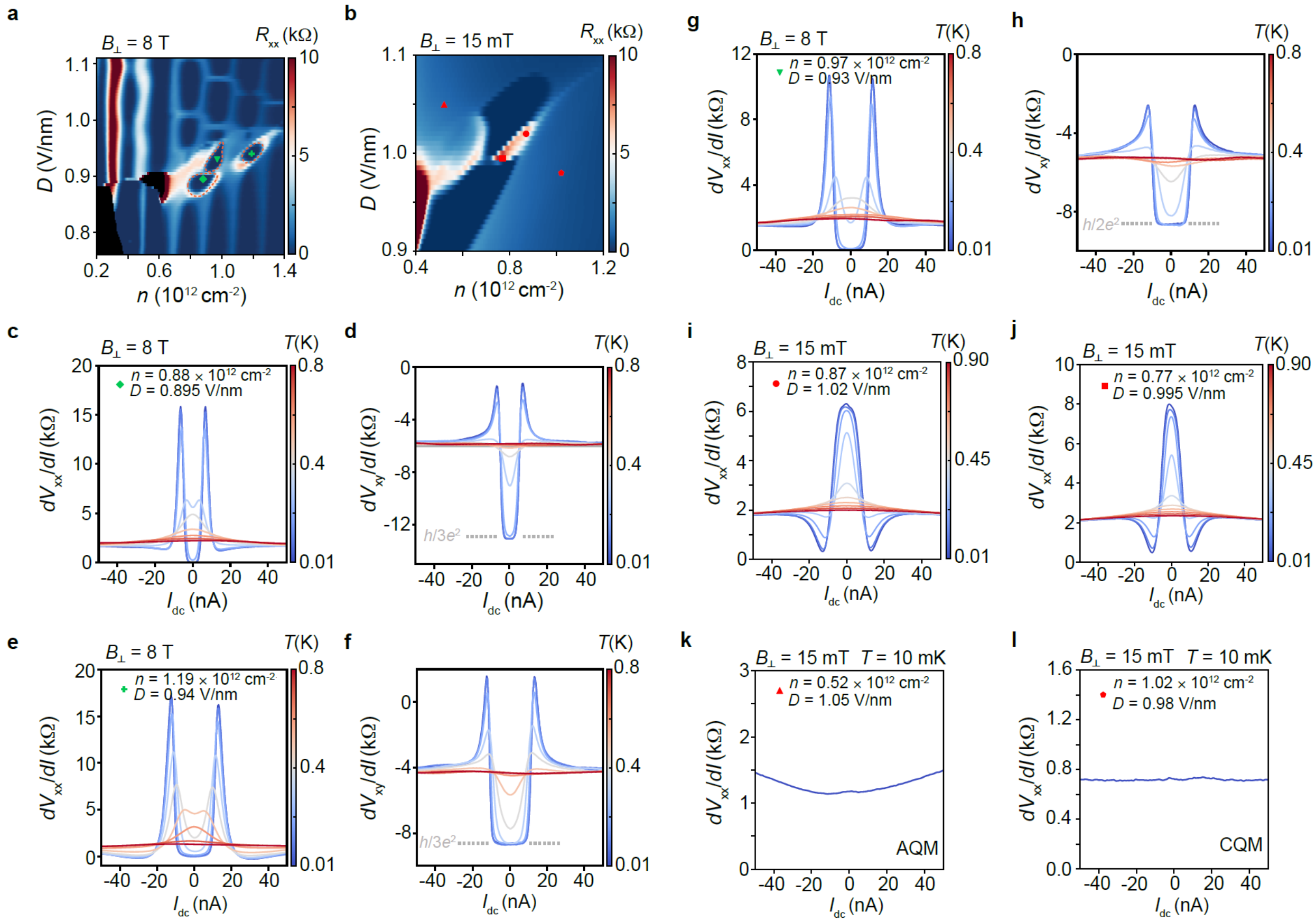

**Extended Data Fig. 10 | Differential resistance measurements near the Lifshitz-transition boundary. a, b,** $R_{xx}$ maps as functions of carrier density $n$ and displacement field $D$, measured at $T$ = 10 mK under $B_\perp$ = 8 T (a) and 15 mT (b). **c-h,** Temperature dependence of the differential resistances $dV_{xx}/dI_{dc}$ and $dV_{xy}/dI_{dc}$ as functions of dc bias current $I_{dc}$ at $B_\perp$ = 8 T, measured at representative points within the RQH states marked in **a**. The observed threshold-like nonlinear response is consistent with depinning behavior, suggesting that electronic crystal may develop in these RQH states. **i, j,** Temperature dependence of $dV_{xx}/dI_{dc}$ as a function of $I_{dc}$ at $B_\perp$ = 15 mT, near the Lifshitz-transition boundary indicated in **b**. **k, l,** $dV_{xx}/dI_{dc}$ vs $I_{dc}$, measured at representative points in the AQM and CQM regions at $T$ = 10 mK and $B_\perp$ = 15 mT, showing no pronounced nonlinear response. These observations suggest that the threshold-like nonlinear response may be associated with the emergence of an electronic crystal near the Lifshitz-transition boundary, whereas no comparable nonlinear behavior is observed in the AQM or CQM regions.

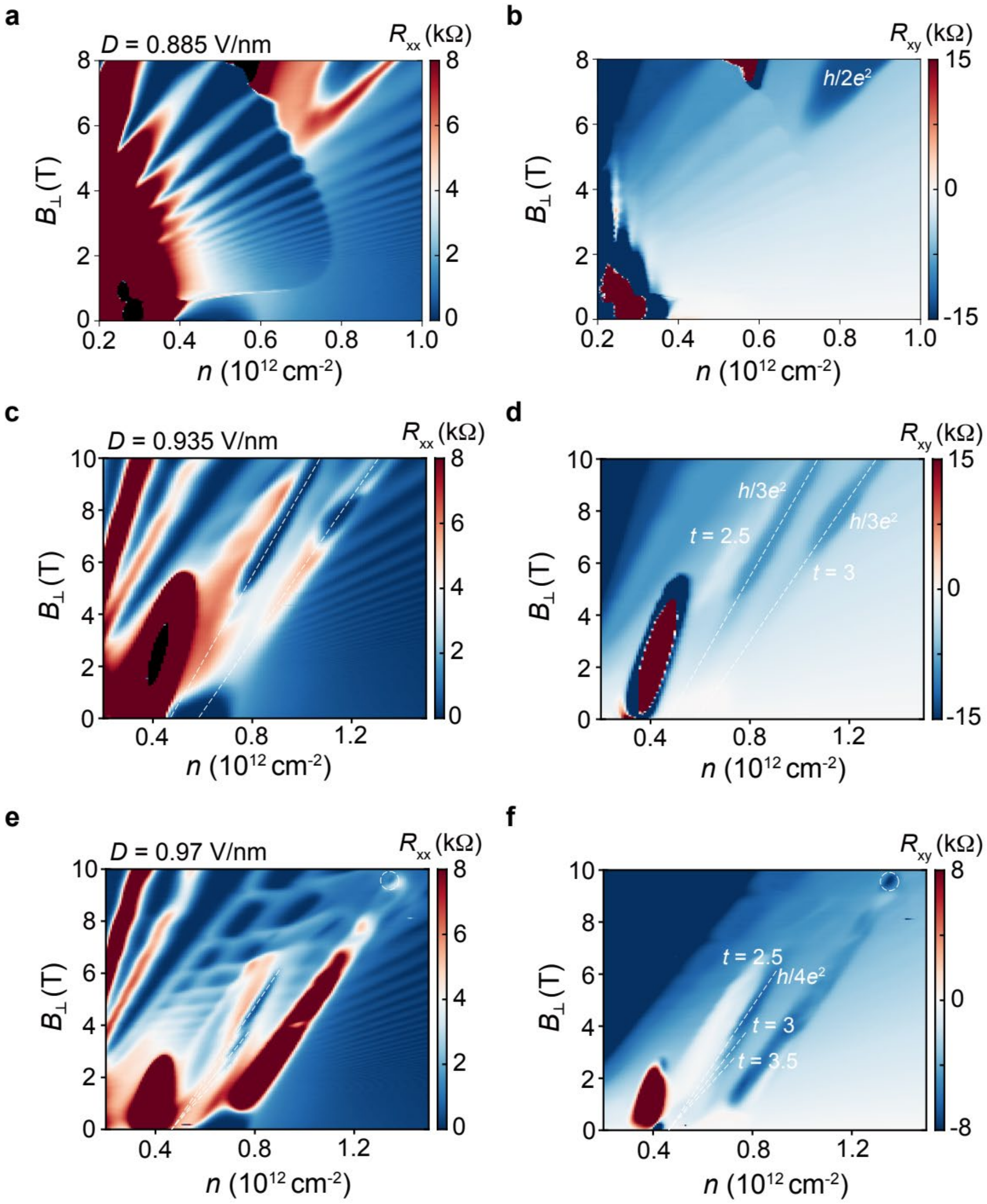

**Extended Data Fig. 11 | Evolution of RQH states with displacement field. a-f,** Landau fan diagrams of $R_{xx}$ and $R_{xy}$ as functions of carrier density $n$ and perpendicular magnetic field $B_\perp$, measured at fixed displacement fields $D = 0.885$ V/nm (a,b), 0.935 V/nm (c,d) and 0.97 V/nm (e,f). RQH states are observed near the Lifshitz-transition boundary. Their associated quantized Hall responses evolve with both $B_\perp$ and $D$, giving rise to different Chern numbers in different field regimes. In **c-f**, RQH states trajectories are marked by dashed lines together with the corresponding quantized Hall resistance values. The slopes of some RQH trajectories do not correspond to the Chern numbers inferred from the quantized $R_{xy}$, indicating that these states do not obey the conventional Středa relation.

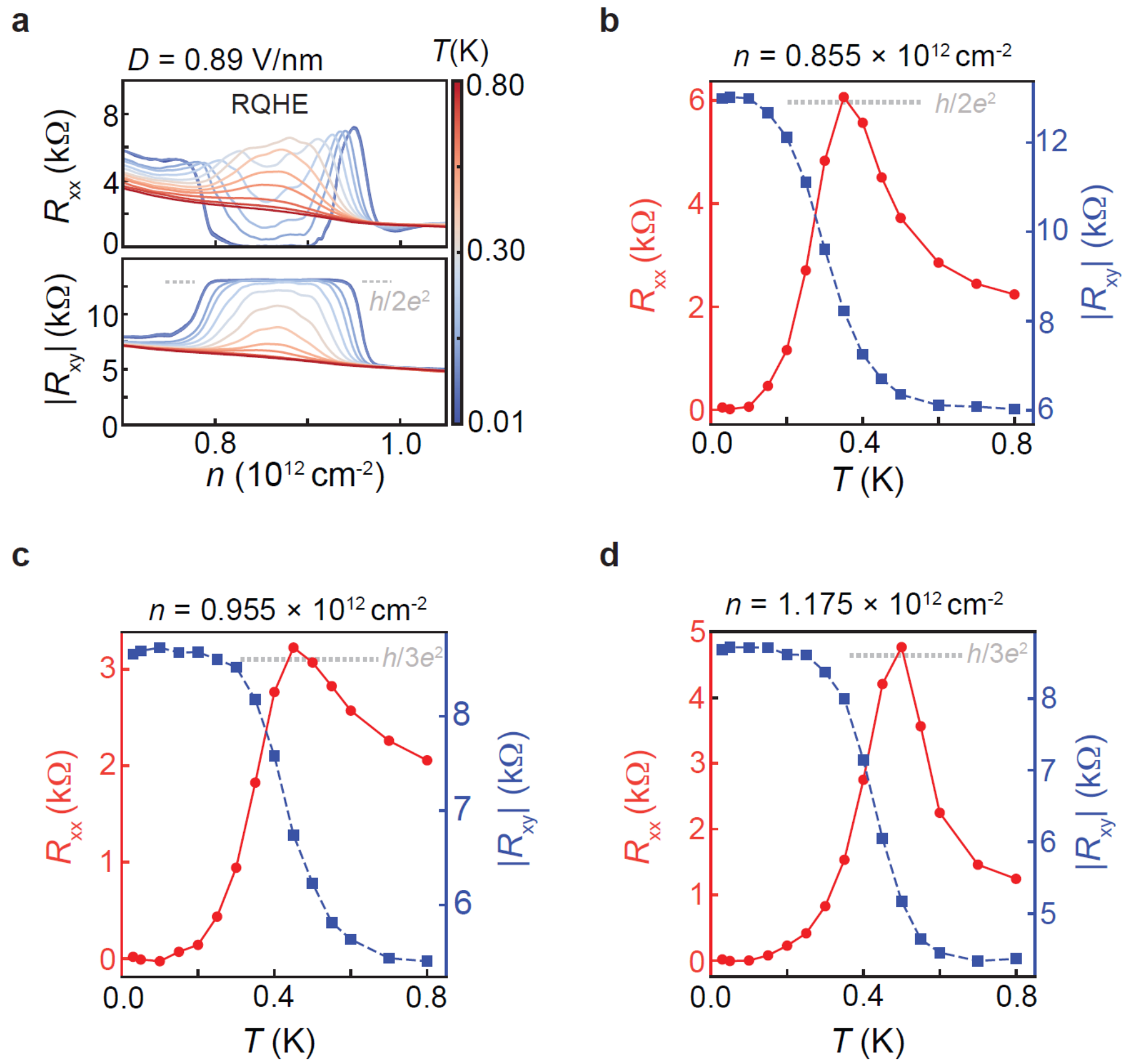

**Extended Data Fig. 12 | Temperature dependence of RQH states. a**, Temperature dependence of $R_{xx}$ and $R_{xy}$ as functions of carrier density $n$ at $D = 0.89$ V/nm, measured near the $C = 2$ RQH state. **b-d**, Temperature dependence of $R_{xx}$ and $R_{xy}$, measured at representative points within the RQH states with $C = 2$ (b), $C = 3$ (c,d).